\documentclass[aip,jcp,amsmath,amssymb,preprint]{revtex4-2}
\usepackage{graphicx} 
\usepackage{bm} 
\usepackage{siunitx} 
\usepackage[version=4]{mhchem} 

\newcommand{\acnote}[1]{\par\noindent
  \parbox{\linewidth}{\footnotesize
    \leftskip=0pt \rightskip=0pt \parfillskip=0pt plus 1fil
    \noindent #1}}

\begin{document}

\title{Development of a Non-Empirical Exchange-Hole Dipole Moment Dispersion Model}

\author{Alastair J. A. Price} \email{alastairj.price@utoronto.ca}
\affiliation{Department of Chemistry, University of Toronto, St. George campus, Toronto, ON, Canada}
\affiliation{Laboratory for AI and Automation, Acceleration Consortium, University of Toronto, 80 St George St, Toronto, ON M5S 3H6, Canada}
\affiliation{Vector Institute for Artificial Intelligence, Toronto, ON M5S 1M1, Canada}

\author{Alberto Otero-de-la-Roza} \email{oteroalberto@uniovi.es}
\affiliation{Departamento de Química Física y Analítica and MALTA
  Consolider Team, Facultad de Química, Universidad de Oviedo,
  33006 Oviedo, Spain}

\date{\today}

\begin{abstract}

The inclusion of dispersion effects is important in density-functional
theory (DFT) to model non-covalent interactions correctly, a task that
is essential in many applications of the theory. Many dispersion
functionals have been proposed in the past. The exchange-hole dipole
moment (XDM) model combines the simplicity of a damped pairwise
asymptotic expression for the dispersion energy with a theory-grounded
approach to calculate the dispersion coefficients. XDM is, arguably,
the most accurate dispersion correction for the description of
molecular crystals and it has been thoroughly tested for other
applications across a wide range of chemistries. Here, we
address the two main shortcomings of XDM. First, XDM relies on the use
of experimentally determined free-atom polarizabilities. Second,
because the XDM atom-in-molecule properties (volumes,
polarizabilities, exchange-hole dipole moments) use the Hirshfeld
partition method, XDM describes systems with large atomic partial
charges, like alkali cations or halide anions, poorly. We propose
neXDM, a non-empirical variant of XDM that removes the experimental
parameters by using the Kirkwood polarizability formula, thereby
making neXDM a pure meta-GGA dispersion functional. In addition,
following previous work by Bu\v{c}ko et al.\ on the similar
Tkatchenko--Scheffler (TS) method, we replace the Hirshfeld
partitioning with its iterative counterpart. The performance of neXDM
is shown to be on par with XDM in standard molecular and crystal
benchmark sets, and greatly improves the modeling of ionic
systems. The new neXDM method sets a new record for the best
dispersion-corrected generalized-gradient approximation (GGA)
functional for molecular crystal lattice energies in the X23 set
(0.700~kcal/mol).

\end{abstract}

 \keywords{}

\maketitle

\section{Introduction}
\label{s:intro}

Non-covalent interactions (NCIs) determine the structure and
energetics of many important systems: molecular crystals, molecules
adsorbed on surfaces, layered materials, large biological molecules,
etc.\cite{grimme2016,hermann2017,stohr2019,beran2016,hoja2019,goerigk2017}
Density-functional theory (DFT) is the main method for the
computational description of these systems, but semilocal and hybrid
exchange-correlation functionals do not capture dispersion effects,
and therefore require a dispersion correction to be
usable.\cite{grimme2016,hermann2017,stohr2019} Many such corrections
have been proposed over the past two decades. Non-local correlation
functionals incorporate dispersion directly into the
self-consistent-field (SCF) calculation of the correlation
energy,\cite{dion2004,vydrov2010b} whereas other approaches use
post-SCF terms that add a damped asymptotic pairwise or
many-body expansion dispersion term to the DFT
energy.\cite{grimme2010,caldeweyher2019,tkatchenko2009,tkatchenko2012,ambrosetti2014,steinmann2011,becke2005,becke2007}
When combined with an appropriate base functional, dispersion
corrections routinely achieve good accuracy for intermolecular binding
energies and molecular crystal lattice
energies.\cite{grimme2016,hermann2017,price2023}

The exchange-hole dipole moment (XDM)
model\cite{becke2005,becke2007,johnson2006,johnson2017} is a post-SCF
dispersion correction. In XDM, the dispersion energy arises from the
interaction between the instantaneous dipoles formed by electrons and
their exchange holes, which are evaluated using the Becke--Roussel
hole model\cite{becke1989} from the electron density, its derivatives,
and the kinetic energy density. Consequently, XDM is formally a
meta-GGA functional: the atomic exchange-hole multipole moments, and
from them the $C_6$, $C_8$, and $C_{10}$ pairwise dispersion
coefficients, follow from the self-consistent
wavefunction.\cite{becke2007,kannemann2010,price2023} This physical
grounding translates into an excellent and remarkably uniform
performance across a wide range of chemistries including gas-phase
dimers and molecular thermochemistry,\cite{xdmhybrid} molecular
crystals, for which XDM-corrected hybrid functionals give the lowest
lattice-energy errors reported to date for the X23 and ICE13
sets,\cite{oterodelaroza2012,price2023,price2023b,mayo2024} surface
adsorption,\cite{christian2016} and layered
materials.\cite{oterodelaroza2020b} Two ingredients of XDM are not
derived from the self-consistent density: the atom-in-molecule
polarizabilities that enter the dispersion coefficients are obtained
by scaling tabulated free-atom polarizabilities with the ratio of the
atom-in-molecule to the free-atom
volume,\cite{becke2007,tkatchenko2009} and both the volumes and the
exchange-hole moments are distributed over the atoms using the
Hirshfeld partitioning of the electron density.\cite{hirshfeld1977}

The Hirshfeld partitioning method assigns to each atom the fraction of
the promolecular density contributed by the corresponding neutral free
atom, an arbitrary choice of reference that ignores the actual partial
charge of the atom within the
system.\cite{bultinck2007,bultinck2007b,parr2005} This bias towards
the neutral free-atom
reference,\cite{ayers2000,nalewajski2000,ayers2006} results in atomic
charges that are systematically too small (by roughly a factor of
three\cite{bultinck2007}), and in the alkali halides they are only
$\pm 0.2$~e against a nominal valence of
unity.\cite{bucko2013b,bucko2014} This has qualitative consequences
for the volume-scaled polarizabilities: for instance, cations are
predicted to be more polarizable than the isoelectronic anions, which
in turn distorts the dispersion coefficients calculated from
them.\cite{bucko2013b,bucko2014} This deficiency impacts the
calculated properties of ionic molecular clusters and
ionic solids in dispersion corrections that use the Hirshfeld
method.\cite{bucko2013,kim2020,caldeweyher2020,nickerson2023}

Within the TS family of methods, several remedies have been
proposed. Bu\v{c}ko \emph{et al.}\cite{bucko2013b,bucko2014} replaced
the Hirshfeld weights by those of the iterative Hirshfeld (HI)
scheme,\cite{bultinck2007,vanpoucke2013} in which the atomic reference
densities are made self-consistent with the atomic populations. The
authors showed that the resulting TS/HI method corrects the lattice
constants, bulk moduli, and atomization energies of the alkali
halides. Gould \emph{et al.}\cite{gould2016,gould2016b} proposed the
fractionally ionic (FI) variant of the many-body dispersion (MBD)
method, in which the reference polarizability and volume are those of
a free partially charged ion evaluated at the HI electron
number.\cite{gould2016} The D4
model\cite{caldeweyher2019,caldeweyher2020} scales its reference
polarizabilities with partial charges from electronegativity
equilibration.

In this work, we show that, because of its reliance on the Hirshfeld
partitioning and the volume-scaled free-atom polarizabilities, XDM is
affected similarly for ionic systems by the aforementioned
drawbacks. While we could consider approaches similar to FI-MBD, it is
not clear \emph{a priori} that these solutions transfer readily from
TS and MBD to XDM, since the way in which the static polarizability
enters the calculation of the dispersion energy in XDM is different
from TS and MBD. In addition, we find that the use of tabulated atomic
and ionic polarizabilities,\cite{gould2016b} the former already
present in XDM, is somewhat at odds with the otherwise non-empirical
character of the model. Removing the free-atom polarizabilities from
XDM has the additional side benefit of dropping the use of linear
scaling of the atomic polarizabilities with volume, which is itself a
dubious
approximation.\cite{gould2016c,manz2019,szabo2022,goger2024,rehman2026}

Based on these considerations, we propose a non-empirical variant of
XDM, termed neXDM, in which we replace the Hirshfeld partition of the
volumes and exchange-hole moments with its iterative
counterpart. However, unlike TS/HI or FI-MBD, we replace the
volume-scaled polarizabilities with an atom-in-molecule polarizability
functional based on the Kirkwood
formula,\cite{kirkwood1932,vinti1932,buckingham1937} which, although
not perfect in terms of accuracy for atoms and
molecules,\cite{goger2024} does capture the correct dependence on the
chemical environment and the scaling with system
size.\cite{szabo2022,goger2024} These two changes make neXDM a pure
meta-GGA dispersion functional. We assess the newly developed neXDM
method on the molecular polarizabilities and dispersion coefficients,
and on molecular and solid-state benchmark sets.  For neutral
molecules and molecular crystals, neXDM performs similarly to XDM. For
systems with charged atoms, the improvement is dramatic: the mean
absolute deviation of the alkali-halide polarizabilities falls from
201 to 11~bohr$^3$, the error in the bulk moduli from 9.3 to 0.9~GPa,
and the circa 30\%\ overbinding of the layered dichalcogenide
materials is essentially removed. We believe neXDM is an important
step towards a dispersion correction that is accurate, non-empirical,
physics-based, and truly universal.

\section{Methodology}
\label{ss:methodology}

\subsection{Iterative Hirshfeld Atomic Partitioning}
\label{ss:hi}

The XDM model uses distributed atom-in-molecule dispersion
coefficients to calculate the dispersion energy via a damped
asymptotic expression (Sec.~\ref{ss:nexdm}). Obtaining the dispersion
coefficients ($C_n$) in XDM requires the atomic partitioning of a
number of molecular properties. For this purpose, XDM uses the
Hirshfeld partitioning method:\cite{hirshfeld1977}
\begin{equation}
  P_A = \int p(\mathbf{r}) w_A^{0}(\mathbf{r}) d^3\mathbf{r}
\end{equation}
where $p$ is some property density, $P_A$ is the integrated property
for atom A, and the Hirshfeld weight is defined as:
\begin{equation}
  w^0_A(\mathbf{r}) = \frac{n_A^0(\mathbf{r})}{\sum_B n_B^0(\mathbf{r})}
\end{equation}
where $n_A^0$ is the in-vacuo atomic density for atom $A$. The atomic
densities in the Hirshfeld method are fixed and correspond to neutral
atoms regardless of the actual atom-in-molecule charge. This
arbitrary choice of reference\cite{bultinck2007,bultinck2007b} causes the
Hirshfeld charges to be smaller than those obtained from other
approaches.\cite{bultinck2007,vanpoucke2013,bucko2014}

The iterative Hirshfeld (HI) method\cite{bultinck2007} removes this
arbitrariness in the choice of reference by requiring that the density
of each atom be consistent with its electron population. The HI
weights are built from a set of atomic references $n_A^{i}$, starting
with the neutral densities and updated until self-consistency. In each
step, the weight at iteration $i$ is calculated as:
\begin{equation}
  \label{eq:hiw}
  w_A^{i}(\mathbf{r}) = \frac{n_A^{i}(\mathbf{r})}{\sum_B n_B^{i}(\mathbf{r})}
\end{equation}
then the atomic populations are obtained:
\begin{equation}
  \label{eq:hin}
  N_A^{i+1} = \int n(\mathbf{r}) w_A^{i}(\mathbf{r}) d^3\mathbf{r}
\end{equation}
and finally the reference densities are rebuilt for the new
populations by linear interpolation between the two integer charge
states that bracket them:
\begin{equation}
  \label{eq:hiref}
  n_A^{i+1}(\mathbf{r}) = n_A^{\lfloor N_A^{i+1}\rfloor}(\mathbf{r})
  \times \xi + n_A^{\lceil N_A^{i+1}\rceil}(\mathbf{r}) \times (1-\xi)
\end{equation}
where $\xi = \lceil N_A^{i+1} \rceil - N_A^{i+1}$ and $n_A^{N}$ are
the reference densities (Sec.~\ref{ss:refdens}). The converged HI
densities and weights are independent of the starting
guess,\cite{bultinck2007,vanpoucke2013} and the resulting atomic
charges correlate considerably better with those derived from the
electrostatic potential than plain Hirshfeld.\cite{vanpoucke2013}

The linear interpolation in Eq.~\ref{eq:hiref} is consistent with the
known behavior of the exact electron density for an open system with
a fractional number of electrons,\cite{perdew1982} and this expression
means that the densities bracketing the usual atomic partial charges
are required, which may be problematic for anions (see
Sec.~\ref{ss:refdens}). The Hirshfeld partition and the HI method can
be derived from information theory as the partitions that minimize the
information loss of the atomic densities relative to their
references.\cite{nalewajski2000,ayers2006}

\subsection{Reference Atomic Densities}
\label{ss:refdens}

The use of the iterative Hirshfeld method requires a set of in-vacuo
electron densities for the neutral atoms as well as the cations and
anions that bracket the atom-in-molecule electron populations
($n_Z^{N}(r)$, with $Z$ the atomic number and $N$ the electron
number). To generate them, we modified the \texttt{atomic} code from
the Quantum ESPRESSO 8.0
distribution,\cite{giannozzi2009,giannozzi2017} which solves the
radial Kohn--Sham equations for a single atom. A comprehensive list of
electron densities of all elements up to $Z = 118$ (neutral, all +1
cations and -1 anions, and all chemically relevant ions with higher
charges) was obtained, using ground-state configurations from the
literature where
available.\cite{kramida2024,andersen1999,rienstrakiracofe2002} The
calculations were carried out spin-restricted, using the B86bPBE
functional,\cite{b86b,pbe} and in the scalar relativistic
approach.\cite{koelling1977}

Neutral atoms and cations were run with the unmodified in-vacuo atomic
Hamiltonian with the mesh extending to 100~bohr. Anions are unbound
using semilocal functionals,\cite{vanpoucke2013,bucko2014} which
precludes SCF convergence.\cite{bucko2013b,bucko2014} To address this
problem, we use an approach similar to that of Gould and
Bu\v{c}ko\cite{gould2016b}: We calculated the electron density of the
anions in the frozen potential of the corresponding neutral atoms. For
the closed-shell halide monoanions, which are experimentally
observable bound species, we use a confinement method to stabilize the
SCF. This is achieved simply by having the radial mesh terminate at
$3.6$ times the radius enclosing 99\%\ of the electrons of the
corresponding neutral atom, which corresponds to assuming the
potential is flat beyond the last point in the radial mesh. The
confinement approach is also used for anions where the last electron
is not bound, for instance, when it enters an empty subshell.

The procedure described above results in 743 atomic densities,
tabulated on radial meshes. To simplify the use of these densities in
electronic structure codes, we fit them using a linear combination of
Slater-type radial functions\cite{clementi1974,koga1997b,koga1999}:
\begin{equation}
  \label{eq:rhofit}
  n_{\text{fit}}(r) = \sum_i c_i\, r^{n_i} \text{e}^{-\alpha_i r}
\end{equation}
where $n_i$ are non-negative integers and $\alpha_i$ are positive
exponents. The coefficients are obtained by minimizing the
electrostatic self-energy of the residual
density\cite{baerends1973,dunlap1979,koster2003}:
\begin{equation}
  \label{eq:eself}
  E_{\text{self}} = \tfrac{1}{2}
\int\!\!\int
\frac{\delta n(\bm{r}) \delta n(\bm{r}')}
{|\bm{r}-\bm{r}'|}\, \text{d}^3\bm{r}\,\text{d}^3\bm{r}'
\end{equation}
with $\delta n(\bm{r}) = n(\bm{r}) - n_{\text{fit}}(\bm{r})$ and
$n(\bm{r})$ represents the numerical density obtained from the atomic
SCF calculation. This is a linear least-squares problem, and other
functionals for the minimization are possible.\cite{verstraelen2016}
Three constraints are imposed on the fits: The $c_i$ coefficients must
be non-negative, the fitted density must integrate to the exact
electron count, and it must also reproduce the third moment of the
numerical density ($4\pi \int_0^\infty n(r) r^5\,\text{d}r$, the free
volume in XDM and TS).

To choose the best $n_i$ and $\alpha$ exponents for a particular ion
($n_Z^{N}(r)$), we precalculate the density contributions from all
$n_i = 0,\dots, 10$ and all $\alpha$ between $0.2$ and $200Z$ in an
even-tempered series with ratio 1.1. The least-squares fit is
carried out using the $\ell_1$ regularization
(LASSO),\cite{tibshirani1996} which performs variable selection,
resulting in a lean analytical representation of the numerical density
for an appropriate choice of the LASSO penalty. This penalty is chosen
to make the integrated absolute difference between the fitted and
reference densities ($4\pi \int_0^{\infty} \left| n(r) -
n_{\text{fit}}(r) \right| r^2\,\text{d}r$) less than $10^{-3}Z$
electrons. Because of the confinement effect on the density, this
accuracy criterion cannot be achieved for some monoanions whose extra
electron enters an empty subshell (e.g.\ \ce{He-}) and some
lanthanide and actinide monoanions; these will rarely be used in
practice. For these atoms, we chose the smallest LASSO penalty that
yields an equivalent number of terms as for the corresponding neutral
atom. The numerical electron densities, coefficients and exponents of
the analytical densities in Eq.~\ref{eq:rhofit}, and the corresponding
comparison plots, are given in the supplementary material (SI).

\subsection{Non-empirical XDM}
\label{ss:nexdm}

In the XDM model,\cite{becke2005,becke2007} the dispersion energy is
added to the energy of a base density functional:
\begin{equation}
  \label{eq:etot}
  E = E_{\text{DFT}} + E_{\text{disp}}
\end{equation}
where $E_{\text{disp}}$ is calculated from atomic dispersion
coefficients that are themselves derived from the system's
wavefunction. The XDM dispersion energy is modeled as arising from the
interaction of the dipoles on different atoms, formed by electrons and
their exchange holes. In the canonical implementation, XDM does not
evaluate the exact exchange hole, which would require the full
one-electron density matrix, but the spherically averaged hole in the
Becke--Roussel (BR) model,\cite{becke1989} consisting of a single
exponential a distance $b_\sigma$ from the reference electron. The
three parameters in the BR model are determined by the known
exact properties of the exchange hole (normalization, depth, and
curvature at the reference point). The BR hole model turns XDM into a
meta-GGA dispersion functional that requires only the local density,
its derivatives, and the kinetic energy density for its computation.

Assuming the hole dipole points towards the nearest nucleus, the
atomic exchange-hole multipole moments are:
\begin{equation}
  \label{eq:moments}
  \langle M_l^2 \rangle_A = \sum_\sigma \int w_A(\mathbf{r})\,
  n_\sigma(\mathbf{r}) \left[ r_A^l - (r_A - b_\sigma)^l \right]^2
  \text{d}^3\mathbf{r}
\end{equation}
with $l = 1$, $2$, and $3$, where $r_A$ is the distance to nucleus
$A$, and $w_A$ is the atomic weight (Sec.~\ref{ss:hi}). The same
weights define the atomic populations, volumes, and second moments:
\begin{equation}
  \label{eq:atprops}
  \begin{aligned}
    N_A &= \int w_A(\mathbf{r})\, n(\mathbf{r})\, \text{d}^3\mathbf{r} \\
    V_A &= \int w_A(\mathbf{r})\, n(\mathbf{r})\, r_A^3\, \text{d}^3\mathbf{r} \\
    \langle r^2 \rangle_A &= \int w_A(\mathbf{r})\, n(\mathbf{r})\, r_A^2\,
      \text{d}^3\mathbf{r}
  \end{aligned}
\end{equation}
The pairwise dispersion coefficients follow from second-order perturbation
theory:
\begin{align}
  \label{eq:c6}
  C_{6,AB} &= F_{AB}\, \langle M_1^2\rangle_A \langle M_1^2\rangle_B \\
  \label{eq:c8}
  C_{8,AB} &= \tfrac{3}{2}\, F_{AB} \left( \langle M_1^2\rangle_A \langle M_2^2\rangle_B
              + \langle M_2^2\rangle_A \langle M_1^2\rangle_B \right) \\
  \label{eq:c10}
  C_{10,AB} &= 2\, F_{AB} \left( \langle M_1^2\rangle_A \langle M_3^2\rangle_B
              + \langle M_3^2\rangle_A \langle M_1^2\rangle_B \right) \nonumber \\
            &\quad + \tfrac{21}{5}\, F_{AB}\, \langle M_2^2\rangle_A \langle M_2^2\rangle_B
\end{align}
with:
\begin{equation}
  \label{eq:fab}
  F_{AB} = \frac{\alpha_A \alpha_B}
                {\langle M_1^2\rangle_A \alpha_B + \langle M_1^2\rangle_B \alpha_A}
\end{equation}
where $\alpha_A$ is the atom-in-molecule polarizability. Unlike
the TS and MBD methods,\cite{tkatchenko2009} the polarizabilities in
XDM are static and do not by themselves yield the dispersion
coefficients: the frequency dependence of the response is carried by
the moments, and $\alpha_A$ enters only through $F_{AB}$.

The atomic polarizabilities in XDM are calculated by scaling the
free-atom values with the atomic volumes:
\begin{equation}
  \label{eq:alpha0}
  \alpha_A = \frac{V_A}{V_A^{0}}\, \alpha_A^{0}
\end{equation}
where $V_A^{0}$ and $\alpha_A^{0}$ are the volume and polarizability
of the free atom, the latter obtained from
experiment.\cite{schwerdtfeger2019,schwerdtfeger2019b,crc88} The same scaling
is used in the later TS method.\cite{tkatchenko2009} Besides the fact
that they introduce empirical parameters into an otherwise physically
grounded dispersion functional, there are two additional difficulties
with Eq.~\ref{eq:alpha0}. First, as with the Hirshfeld weights,
the free-atom polarizabilities correspond to the neutral atom,
regardless of the atom-in-molecule partial charge in the actual
system. In the MBD@rsSCS method, using charge-dependent
polarizabilities has been shown to greatly improve the description of
ionic systems.\cite{gould2016} Second, the actual scaling of atomic
polarizabilities with volume has been shown to deviate substantially
from linearity.\cite{gould2016c,manz2019,szabo2022,rehman2026}

The non-empirical XDM model (neXDM) we propose is based on two changes
to the method described above. First, we replace the Hirshfeld atomic
partition (the weights in Eqs.~\ref{eq:moments} and~\ref{eq:atprops})
with its iterative Hirshfeld counterpart. Second, the atomic
polarizability is computed using the Kirkwood
formula:\cite{kirkwood1932,vinti1932}
\begin{equation}
  \label{eq:alphak}
  \alpha_A = \frac{4}{9}\, \frac{\langle r^2 \rangle_A^2}{N_A}
\end{equation}
which is based on a variational estimate of the second-order
perturbation theory expression for the polarizability, together with a
simplifying assumption that allows it to be obtained from the density
alone. While the Kirkwood atomic polarizabilities are not
accurate,\cite{goger2024} they do capture the correct scaling with
system size and the response of $\alpha$ to changes in the atomic
chemical environment, which, as we shall see, results in accurate
dispersion coefficients (Sec.~\ref{ss:alphac6}). The use of the
Kirkwood polarizability removes the dependence of the
dispersion model on the experimental static polarizabilities.

Once the dispersion coefficients are obtained, the calculation of the
dispersion energy proceeds in the same way as in XDM,
namely\cite{johnson2006,becke2007}:
\begin{equation}
  \label{eq:edisp}
  E_{\text{disp}} = -\frac{1}{2} \sum_{A \neq B} \sum_{n=6,8,10}
  \frac{C_{n,AB}}{R_{\text{vdw},AB}^n + R_{AB}^n}
\end{equation}
where $R_{AB}$ is the interatomic distance and the sum of van der Waals
radii is:
\begin{equation}
  \label{eq:rvdw}
  R_{\text{vdw},AB} = a_1 R_{c,AB} + a_2
\end{equation}
with the critical radius $R_{c,AB}$ being the average of the three
crossover distances $(C_{8}/C_{6})^{1/2}$, $(C_{10}/C_{6})^{1/4}$ and
$(C_{10}/C_{8})^{1/2}$ at which the corresponding terms contributing
to the dispersion energy become equal. The $a_1$ and $a_2$ parameters
are fitted for a particular functional and basis set against a small
set of reference dimer energies (Sec.~\ref{ss:parametrization}).

\section{Computational Details}
\label{s:compdetails}

The neXDM method was implemented in three programs: postg, a
stand-alone program that evaluates the XDM dispersion energy from a
converged molecular wavefunction,\cite{xdmhybrid} the
numerical atomic orbital code FHI-aims,\cite{blum2009} and the
plane-wave Quantum ESPRESSO package.\cite{giannozzi2009,giannozzi2017}
The iterative Hirshfeld populations are converged by fixed-point
iteration of Eqs.~\eqref{eq:hiw}--\eqref{eq:hiref}, until the atomic
populations are converged to $10^{-7}$ electrons. In geometry
relaxations, the converged populations of the previous step are used
as the starting guess for the next. For the reference densities, we
used the analytic expressions of Sec.~\ref{ss:refdens}.

The Gaussian~16 program\cite{g16} was used for the molecular
calculations using linear combinations of atomic orbitals (LCAO)
expressed as Gaussian type orbitals (GTO). These calculations were run
in combination with the reasonably large aug-cc-pVTZ basis set (except
where noted) and on an ultrafine integration grid. We used this basis
set because it provides a nice balance between size and cost, contains
diffuse functions, which are necessary to model NCI,\cite{xdmbasis}
and also for consistency with our previous works.\cite{xdmhybrid} The
aug-cc-pVTZ basis set does not provide bases for some heavy elements
in the examined benchmark sets (e.g.\ GMTKN55). Although a basis set
that covers more of the periodic table could have been used, we
decided against using pseudopotentials since the pseudo-densities may
have an impact when used in the XDM equations, and we wanted to avoid
compounding errors in our analysis. For the LCAO calculations, sixteen
functionals were coupled with XDM for the molecular (Gaussian+postg)
calculations: the GGA functionals BLYP,\cite{becke1988b,lee1988}
BP86,\cite{becke1988b,perdew1986} BPBE,\cite{becke1988b,pbe}
PBE,\cite{pbe} and PW86PBE;\cite{perdew1986b,pbe,kannemann2010} the
meta-GGA TPSS;\cite{tao2003} the global hybrids
B3LYP,\cite{becke1993,stephens1994} B3P86,\cite{becke1993,perdew1986}
B3PW91,\cite{becke1993,perdew1992} B97-1,\cite{hamprecht1998}
BHandHLYP,\cite{becke1993b,lee1988} and
PBE0;\cite{adamo1999,ernzerhof1999} the range-separated hybrids
CAM-B3LYP,\cite{yanai2004} HSE06,\cite{heyd2003,krukau2006} and
LC-$\omega$PBE;\cite{vydrov2006} and Hartree--Fock (HF).

In FHI-aims, we parametrized the 16 functionals of its built-in XDM
damping table\cite{price2023}: the GGAs B86bPBE,\cite{b86b,pbe}
PBE,\cite{pbe} and revPBE;\cite{zhang1998} the global hybrids B3LYP,
BHLYP, PBE0, revPBE0, and the 25\% and 50\% exact-exchange variants
B86bPBE-25X, B86bPBE-50X, PBE-50X, and revPBE-50X;\cite{price2023} and
the range-separated hybrids HSE06 and
LC-$\omega$PBEh\cite{vydrov2006,rohrdanz2009} with range-separation
parameters of 0.20 and 0.40~bohr$^{-1}$ and 0\% or 20\% short-range
exact exchange. The light, lightdense, lightdenser, intermediate, and
tight basis set plus integration grid tiers were used in combination
with all the listed functionals.

In Quantum ESPRESSO, we parametrized B86bPBE, PW86PBE, BLYP, and
PBE. The projector-augmented-wave (PAW) method\cite{paw} was used with
pseudopotentials from the pslibrary.\cite{pslibrary} Plane-wave
cutoffs were 80 and 800~Ry for the wavefunctions and the density,
respectively.

\section{Results}
\label{s:results}

\subsection{Polarizabilities and Dispersion Coefficients}
\label{ss:alphac6}

Before analyzing the performance of neXDM, it is worth examining how
XDM represents static polarizabilities ($\alpha$) and leading
dispersion coefficients ($C_6$) since these are molecular properties
that are calculable at a high level of theory, and the latter is
responsible for the performance of the dispersion correction as a
whole. We consider two different ways of calculating $\alpha$: i) the
XDM (and TS) scaling of the in-vacuo polarizability with the atomic
volume (Eq.~\ref{eq:alpha0}) and ii) the Kirkwood semilocal formula
(Eq.~\ref{eq:alphak}) entering the neXDM equations.  The models are
referred to in this section as XDM and neXDM. The $C_6$ dispersion
coefficients are calculated using Eq.~\ref{eq:c6}, but in neXDM the
iterative Hirshfeld weights are used for the atomic partitions. The
molecular $\alpha$ and $C_6$ are obtained as a sum of the
corresponding atom-in-molecule atomic values.

\begin{figure*}
\includegraphics[width=\textwidth]{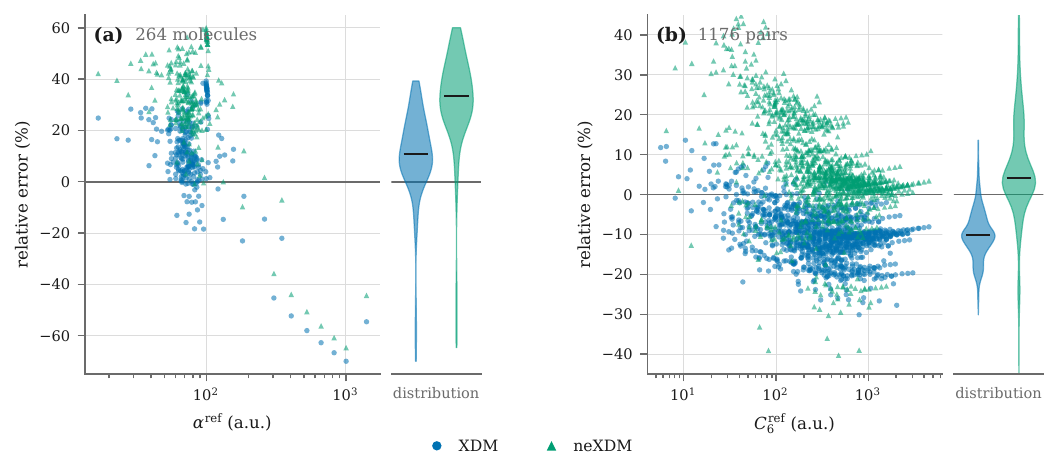}
\caption{Relative errors in the molecular static dipole
  polarizabilities\cite{wilkins2019,yang2019} (a) and in the molecular
  $C_6$ dispersion coefficients\cite{kumar1985,bucko2013b} (b), for
  the XDM and neXDM models at the PBE/aug-cc-pVTZ level, against the
  reference value of each system.}
\label{f:alphac6}
\end{figure*}

Figure~\ref{f:alphac6}(a) shows the molecular polarizabilities for the
264 molecules of the QM7b\cite{montavon2013,rupp2012,blum2009b} and
AlphaML\cite{wilkins2019,yang2019,yang2019b} databases.  Both models
overestimate the polarizabilities, with mean signed and mean absolute
errors of $+10.8$ and 15.8\% for XDM and $+32.0$ and 34.9\% for neXDM
(i.e.\ Kirkwood). The bias is expected of the two models because they
are additive and neglect the effect of electric field screening from
neighboring atoms on the atomic polarizability.\cite{tkatchenko2012}
This effect is very clear for the polarizabilities exceeding 200~a.u.,
corresponding to the polyene, acene, and fullerene series of the
AlphaML set. In these systems, the polarizability grows superlinearly
with the length of the conjugated chain, and cannot be reproduced by a
simple sum of isotropic atomic polarizabilities. However, it is
important to note that the polarizability results in
Figure~\ref{f:alphac6} compound the model errors with the additivity
approximation, and the molecular polarizability itself is never
calculated in either XDM or neXDM.

Figure~\ref{f:alphac6}(b) also shows the $C_6$ coefficients of the
1176 pairs formed by the 48 species for which dipole oscillator
strength distributions are available.\cite{kumar1985,bucko2013b} At
the PBE level, XDM underestimates the coefficients by $-10.1$\% with a
mean absolute error of 10.5\%, and neXDM overestimates them by
$+6.7$\% with a mean absolute error of 10.3\%.  For comparison, the
Tkatchenko--Scheffler method evaluated on these same pairs gives a
mean absolute error of 5.3\%, which degrades to 8.6\% when the plain
Hirshfeld partition is replaced by its iterative
counterpart.\cite{bucko2013b} The neXDM dispersion coefficients are
more accurate for larger molecules with higher dispersion
coefficients. From Figure~\ref{f:alphac6} and the comparison with the
statistics reported for other dispersion corrections, we conclude that
either XDM or neXDM give reasonable molecular dispersion coefficients.

\begin{table*}
\caption{Static dipole polarizabilities ($\alpha$) and dispersion
  coefficients $C_6$ of the alkali cations and the halide anions (in
  a.u.), calculated using the XDM and neXDM models compared with the
  free-standing reference values of Gould and
  Bu\v{c}ko.\cite{gould2016b}}
\label{t:ions}
\begin{tabular}{l@{\hspace{0.6cm}}rrr@{\hspace{0.6cm}}rrr}
\hline\hline
 & \multicolumn{3}{c}{$\alpha$} & \multicolumn{3}{c}{$C_6$} \\
\cline{2-4}\cline{5-7}
 & XDM & neXDM & Ref. & XDM & neXDM & Ref. \\
\hline
Li$^{+}$     & 1.690 & 0.1868 & 0.1930 & 0.7746 & 0.08565 & 0.07909 \\
Na$^{+}$     & 12.29 & 1.919 & 0.9300 & 19.23 & 3.002 & 1.538 \\
K$^{+}$      & 50.61 & 9.358 & 5.050 & 178.3 & 32.96 & 20.96 \\
Rb$^{+}$     & 76.02 & 12.13 & 8.320 & 376.6 & 60.10 & 49.15 \\
Cs$^{+}$     & 124.0 & 21.30 & 15.00 & 864.0 & 148.3 & 129.2 \\
\hline
F$^{-}$      & 10.53 & 16.17 & 15.00 & 67.42 & 103.6 & 73.51 \\
Cl$^{-}$     & 27.32 & 39.91 & 30.30 & 271.7 & 396.8 & 276.5 \\
Br$^{-}$     & 35.86 & 36.95 & 42.82 & 432.4 & 445.6 & 497.3 \\
I$^{-}$      & 58.04 & 51.90 & 61.65 & 871.8 & 779.5 & 925.4 \\
\hline
MAE$^a$          & 28.13 & 4.646 & --- & 151.9 & 43.51 & --- \\
MARE$^a$ (\%)    & 500.2 & 39.07 & --- & 449.3 & 34.26 & --- \\
\hline\hline
\end{tabular}
\acnote{$^a$ Mean absolute error (MAE) and mean absolute relative
  error (MARE)}
\end{table*}

The difficulties of the XDM model for ionic systems are illustrated by
the polarizabilities and $C_6$ coefficients for the alkali cations and
the halide anions shown in Table~\ref{t:ions}. For the cations, XDM
overestimates the polarizability by factors between 8 and 13, and the
dispersion coefficient by comparable amounts. This failure can be
traced back to the neutral-atom reference and the inability of the
volume scaling (Eq.~\ref{eq:alpha0}) to represent changes in the
atomic polarizability due to an increase or decrease in the number of
electrons. Importantly, XDM, same as TS, spuriously predicts cations
being more polarizable than the corresponding isoelectronic
anions,\cite{bucko2014} and the cation $C_6$ are overestimated by an
order of magnitude. The neXDM model, which is simply the Kirkwood
model for the polarizability, is an order of magnitude better for the
same nine ions (39.1\%\ for $\alpha$ and 34.3\%\ for
$C_6$). Therefore, even though in terms of error the Kirkwood model
underpinning neXDM is far from perfect, it is clear that it captures
the correct response to a change in the electron population of a given
atom, and therefore we expect it to perform better than XDM in
molecules and solids with significant interatomic charge transfer. As
we shall see in Sec.~\ref{ss:alkalihalides}, this has a large impact
on the description of ionic solids.

\subsection{Parametrization of neXDM}
\label{ss:parametrization}

\begin{table*}
\caption{Becke--Johnson damping parameters $a_1$ and $a_2$ (in \AA{})
  of neXDM, fitted to the 49 dimers of the KB49 set, and the
  root-mean-square percent deviation RMSP of the fit (\%). The RMSP of
  standard XDM is also indicated.}
\label{t:params}
\scriptsize
\renewcommand{\arraystretch}{0.95}
\begin{tabular*}{\textwidth}{@{\extracolsep{\fill}}lrrrr@{\hspace{1.5em}}lrrrr}
\hline\hline
 & $a_1$ & $a_2$ & RMSP & RMSP &
 & $a_1$ & $a_2$ & RMSP & RMSP \\
 &       &       & neXDM& XDM              &
 &       &       & neXDM& XDM              \\
\hline
\multicolumn{5}{c}{Gaussian~16 / postg, aug-cc-pVTZ} & \multicolumn{5}{c}{FHI-aims, lightdense (cont.)} \\
\hline
BLYP & 0.6293 & 1.4228 & 13.08 & 12.88 & revPBE & 0.8066 & 0.8917 & 15.28 & 16.32 \\
BP86 & 0.6560 & 1.4094 & 25.34 & 25.42 & B3LYP & 0.6438 & 1.7250 & 12.97 & 10.54 \\
BPBE & 0.3487 & 2.0463 & 28.36 & 30.19 & BHLYP$^a$ & 0.0000 & 4.1609 & 15.28 & 13.03 \\
PBE & 0.4000 & 2.8720 & 17.66 & 18.35 & PBE0 & 0.1743 & 3.7778 & 15.89 & 15.33 \\
PW86PBE & 1.0028 & 0.8468 & 14.20 & 14.99 & revPBE0 & 1.0464 & 0.3732 & 12.19 & 10.41 \\
TPSS & 1.0020 & 0.5910 & 12.61 & 13.60 & B86bPBE-25X & 0.8849 & 1.2900 & 14.12 & 12.65 \\
B3LYP & 0.6503 & 1.6020 & 9.31 & 8.28 & B86bPBE-50X & 0.8376 & 1.7188 & 15.44 & 13.72 \\
B3P86 & 0.8637 & 1.0405 & 18.17 & 17.50 & PBE-50X & 0.2440 & 3.8023 & 15.11 & 13.93 \\
B3PW91 & 0.5823 & 1.6077 & 19.82 & 19.13 & revPBE-50X & 1.4289 & -0.4965 & 15.85 & 12.35 \\
B97-1 & 0.0441 & 4.1680 & 16.09 & 16.95 & HSE06 & 0.0861 & 4.0846 & 16.88 & 16.54 \\
BHandHLYP & 0.1533 & 3.4117 & 12.02 & 11.29 & LC-$\omega$PBEh$^b$ 20/00 & 0.6054 & 2.1036 & 13.56 & 12.48 \\
PBE0 & 0.5382 & 2.4386 & 13.33 & 13.57 & LC-$\omega$PBEh$^b$ 20/20 & 0.8390 & 1.4924 & 12.54 & 10.53 \\
CAM-B3LYP & 0.1549 & 3.5640 & 10.37 & 9.89 & LC-$\omega$PBEh$^b$ 40/00 & 1.5458 & -0.7648 & 12.47 & 11.35 \\
HSE06 & 0.4852 & 2.6458 & 14.27 & 14.50 & LC-$\omega$PBEh$^b$ 40/20 & 1.6759 & -1.0181 & 14.38 & 13.44 \\
\cline{6-10}
LC-$\omega$PBE & 1.1674 & 0.2904 & 10.98 & 11.54 & \multicolumn{5}{c}{FHI-aims, tight} \\
\cline{6-10}
HF & 0.1553 & 2.8877 & 16.90 & 19.17 & B86bPBE & 0.9300 & 0.8879 & 13.74 & 14.19 \\
\cline{1-5}
\multicolumn{5}{c}{FHI-aims, light} & PBE & 0.4491 & 2.6723 & 17.53 & 18.12 \\
\cline{1-5}
B86bPBE & 0.4800 & 2.4024 & 16.49 & 16.61 & revPBE & 0.6555 & 1.2758 & 13.81 & 14.41 \\
PBE$^a$ & 0.0000 & 4.1965 & 20.43 & 21.36 & B3LYP & 0.6011 & 1.7211 & 9.37 & 8.06 \\
revPBE & 0.7682 & 0.9929 & 13.80 & 14.98 & BHLYP & 0.1452 & 3.4243 & 12.60 & 11.13 \\
B3LYP & 0.3599 & 2.5977 & 11.30 & 8.71 & PBE0 & 0.5522 & 2.3534 & 12.81 & 12.58 \\
BHLYP$^a$ & 0.0000 & 4.1210 & 14.28 & 11.71 & revPBE0 & 0.7417 & 1.1794 & 9.63 & 9.12 \\
PBE0$^a$ & 0.0000 & 4.2981 & 14.00 & 13.14 & B86bPBE-25X & 0.8639 & 1.1679 & 11.09 & 10.31 \\
revPBE0 & 0.9909 & 0.5246 & 12.25 & 9.40 & B86bPBE-50X & 0.8232 & 1.4032 & 13.40 & 11.88 \\
B86bPBE-25X & 0.6458 & 2.0197 & 12.87 & 10.61 & PBE-50X & 0.6039 & 2.2355 & 13.63 & 12.58 \\
B86bPBE-50X & 0.6130 & 2.3859 & 15.56 & 12.84 & revPBE-50X & 0.8095 & 1.1794 & 11.98 & 9.65 \\
PBE-50X$^a$ & 0.0000 & 4.5258 & 14.67 & 12.97 & HSE06 & 0.5162 & 2.5019 & 13.69 & 13.55 \\
revPBE-50X & 1.3382 & -0.2436 & 17.00 & 11.73 & LC-$\omega$PBEh$^b$ 20/00 & 1.0373 & 0.5979 & 8.83 & 8.62 \\
HSE06$^a$ & 0.0000 & 4.3251 & 14.84 & 14.28 & LC-$\omega$PBEh$^b$ 20/20 & 0.9398 & 0.9684 & 9.23 & 8.01 \\
LC-$\omega$PBEh$^b$ 20/00 & 0.3772 & 2.7989 & 11.89 & 10.75 & LC-$\omega$PBEh$^b$ 40/00 & 1.0477 & 0.6071 & 11.63 & 10.02 \\
LC-$\omega$PBEh$^b$ 20/20 & 0.5527 & 2.3644 & 11.79 & 8.90 & LC-$\omega$PBEh$^b$ 40/20 & 1.0036 & 0.8093 & 12.97 & 11.36 \\
\cline{6-10}
LC-$\omega$PBEh$^b$ 40/00 & 1.4216 & -0.4088 & 13.52 & 11.22 & \multicolumn{5}{c}{Quantum ESPRESSO, PAW} \\
\cline{6-10}
LC-$\omega$PBEh$^b$ 40/20 & 1.3901 & -0.1564 & 15.64 & 13.42 & B86bPBE & 0.6615 & 1.6628 & 13.53 & 14.94 \\
\cline{1-5}
\multicolumn{5}{c}{FHI-aims, lightdense} & PW86PBE & 0.6659 & 1.7836 & 13.85 & 14.79 \\
\cline{1-5}
B86bPBE & 0.6476 & 1.8902 & 18.75 & 18.79 & BLYP & 0.3957 & 2.0421 & 14.43 & 22.02 \\
PBE$^a$ & 0.0000 & 4.2192 & 22.80 & 23.55 & PBE & 0.3198 & 3.0120 & 17.09 & 17.87 \\
\hline\hline
\end{tabular*}

\acnote{$^a$ In cases when the unconstrained fit gave $a_1 < 0$, an
  $a_1 = 0$ constraint was applied.}
\acnote{$^b$ The LC-$\omega$PBEh entries are labeled by the
  range-separation parameter (in bohr$^{-1}$, multiplied by 100) and
  the fraction of short-range exact exchange.}
\end{table*}

The damping parameters $a_1$ and $a_2$ (Eq.~\eqref{eq:rvdw}) were
fitted for neXDM using every combination of code, functional and basis
set (Sec.~\ref{s:compdetails}) against the Kannemann--Becke
set\cite{kannemann2010} (KB49) of 49 gas-phase dimers, following the
same procedure as in previous XDM
parametrizations.\cite{becke2007,xdmhybrid,price2023} The quantity
minimized is the root-mean-square percent deviation (RMSP) from the
reference binding energies. In the few cases in which the
unconstrained fit gave a negative $a_1$, the fit was repeated with
fixed $a_1 = 0$.

Table~\ref{t:params} shows the neXDM damping parameters, together with
the RMSP values obtained from the fit, and the RMSP from the
equivalent fit in XDM for comparison. The neXDM and XDM models show
essentially the same performance and behavior in the parametrization,
with minor differences in RMSP of tenths of a \%. The only significant
differences are: i) neXDM outperforms XDM in combination with
HF/aug-cc-pVTZ, ii) neXDM also outperforms XDM in the Quantum ESPRESSO
fits, particularly in combination with the BLYP functional, and iii)
XDM outperforms neXDM for the hybrid functionals with FHI-aims,
although the difference between the two models decreases as the basis
set increases in size, with relatively small differences at the tight
level.

The complete parameter tables for the neXDM parametrization, including
the remaining FHI-aims tiers not shown in Table~\ref{t:params}, are
given in the SI.

\subsection{Molecular Benchmark Sets}

\begin{table}
\caption{Mean absolute errors (in kcal/mol) of XDM and neXDM on five
  non-covalent interaction benchmark sets, averaged over the sixteen
  functionals of Table~\ref{t:params}, using LCAO and aug-cc-pVTZ.}
\label{t:nci}
\begin{tabular}{lrr}
\hline\hline
 & XDM & neXDM \\
\hline
S22x5\cite{jurecka2006,grafova2010}       & 0.420 & 0.433 \\
S66x8\cite{rezac2011,brauer2016}          & 0.281 & 0.287 \\
IHB100x10\cite{rezac2020}                 & 0.845 & 0.873 \\
SH250x10\cite{kriz2022}                   & 0.779 & 0.795 \\
IONPI19\cite{spicher2021}                 & 1.772 & 1.473 \\
\hline
\multicolumn{3}{c}{Subsets of IONPI19} \\
\hline
Alkali cations             & 1.728 & 0.929 \\
Molecular cations          & 2.451 & 2.537 \\
Halide anions              & 2.213 & 1.985 \\
\ce{NO3-}, \ce{SCN-}       & 1.048 & 1.084 \\
Conformers                 & 0.381 & 0.365 \\
\hline\hline
\end{tabular}
\end{table}

Table~\ref{t:nci} collects the average MAE for XDM and neXDM over
sixteen functionals on five molecular benchmark sets for non-covalent
interactions: S22x5 (110 binding energies of the S22 dimers at five
separations\cite{jurecka2006,grafova2010}), S66x8 (528 binding energies
of the S66 dimers at eight separations\cite{rezac2011,brauer2016}),
IHB100x10 (interaction energies of ionic hydrogen
bonds\cite{rezac2020}), SH250x10 (2050 interaction energies of
halogen-, chalcogen- and pnictogen-bonded complexes\cite{kriz2022}),
and IONPI19 (ion--$\pi$ interaction energies\cite{spicher2021}). The
iodine complexes of SH250x10 and the potassium-bearing reaction of
IONPI19 have been left out because aug-cc-pVTZ does not provide basis
sets for these atoms. The IHB100x10 and IONPI19 sets contain
charged systems. The table with the per-functional MAEs is given in
the SI.

Table~\ref{t:nci} shows that the performance of the neXDM model is
very similar to XDM on four of the five sets. Averaged over the
sixteen functionals, the differences in MAE between neXDM and XDM in
the S22x5, S66x8, IHB100x10, and SH250x10 sets are in the range of a
few hundredths of a kcal/mol, which is smaller than the spread between
functionals within either column and smaller than the uncertainty of
the reference data.  For IONPI19, in contrast, the MAE falls from
1.772 to 1.473~kcal/mol, and neXDM is the better of the two for
thirteen of the sixteen functionals. Table~\ref{t:nci} shows the
partial MAEs for the different types of systems in the IONPI19
set. The decrease in MAE is most notable for the alkali cation--$\pi$
dimers (1.728 to 0.929~kcal/mol) and the halide anion--$\pi$ dimers
(2.213 to 1.985~kcal/mol). For the molecular anions and cations and
the conformational energy differences, the difference between neXDM
and XDM is much smaller, below 0.1~kcal/mol. This is reasonable since
single-atom ions carry higher partial charges than atoms in a
molecular ion, and therefore we expect them to be impacted by the XDM
shortcomings regarding the reference densities. The same argument can
be used to explain the very similar MAEs obtained with XDM and neXDM
for the IHB100x10, for which the partial MAE of the two models agrees
within 0.1~kcal/mol in the cation (0.947 and 0.935~kcal/mol),
monoanion (0.886 and 0.927~kcal/mol), and dianion (0.749 and
0.801~kcal/mol) systems.

\begin{table}
\caption{The WTMAD-2 statistic (in kcal/mol) of XDM and neXDM on the
  GMTKN55 set\cite{goerigk2017} and its subcategories, averaged over
  the sixteen functionals of Table~\ref{t:params}, using LCAO and
  aug-cc-pVTZ.}
\label{t:gmtkn55}
\begin{tabular}{lrr}
\hline\hline
 & XDM & neXDM \\
\hline
Basic                  & 6.38 & 6.18 \\
Large                  & 10.74 & 10.94 \\
Barriers               & 12.57 & 12.43 \\
Inter                  & 5.99 & 5.76 \\
Intra                  & 9.19 & 9.37 \\
All                    & 8.51 & 8.46 \\
\hline\hline
\end{tabular}
\end{table}

Table~\ref{t:gmtkn55} shows the results for XDM and neXDM in the
GMTKN55 dataset,\cite{goerigk2017} composed of 50 subsets (the
original 55 sets minus the ALKBDE10, HEAVYSB11, HEAVY28, HAL59, and
CHB6 sets, which were removed because they contain heavy elements for
which aug-cc-pVTZ does not have a basis set). The table shows the
WTMAD-2 weighted measure\cite{goerigk2017} for each of the categories
in the dataset, as defined in Table~1 of the original
article\cite{goerigk2017}: basic, basic properties and reactions of
small systems; large, reaction energies of large systems and
isomerizations; barriers, reaction barrier heights; inter,
intermolecular non-covalent interactions; intra, intramolecular
non-covalent interactions. The WTMAD-2 is defined as:
\begin{equation}
\text{WTMAD-2} = \frac{1}{\sum_i N_i} \sum_i N_i\, \frac{\overline{|\Delta E|}}{\langle|\Delta E|\rangle_i}\,\text{MAD}_i
\end{equation}
where $i$ runs over the subsets, $N_i$ is the number of data points in
subset $i$, $\langle|\Delta E|\rangle_i$ is its average absolute
reference energy, and $\overline{|\Delta E|} = 56.84$~kcal/mol the
average of the $\langle|\Delta E|\rangle_i$. Within a
category, the WTMAD-2 uses the same expression but with the sums
restricted to the subsets in that category.

The results in Table~\ref{t:gmtkn55} show the same pattern on GMTKN55
as in the non-covalent interaction sets in Table~\ref{t:nci}. Namely,
there is not much difference between the performance of XDM and neXDM,
with variations in WTMAD-2 in the order of one or two tenths of a
kcal/mol.  The complete set of WTMAD-2 for every functional examined
and the MAEs for individual sets are given in the SI. By a large
margin, the single largest change in MAE between XDM and neXDM in the
GMTKN55 database is ALK8, the reactions involving alkali metals, where
the MAE falls from 16.01 to 6.22~kcal/mol when using neXDM instead of
XDM. This decrease in MAE occurs for all sixteen functionals.

As before, the reason for this dramatic improvement can be traced back
to the behavior of XDM for alkali cations. For instance, in the case
of methyllithium, a molecule that enters three data points in ALK8,
the Hirshfeld method (PBE/aug-cc-pVTZ) assigns a partial charge of
0.48~e to the Li atom, compared with 0.76~e with iterative
Hirshfeld. Its XDM atom-in-molecule polarizability is 54.8~bohr$^3$,
more than the rest of the atoms in the molecule combined. In addition
to Li having a lower charge, the reference polarizability for neutral
Li in XDM is 164~bohr$^3$, compared to 0.19~bohr$^3$ for \ce{Li+}. In
contrast, the atom-in-molecule polarizability of Li in methyllithium
is 2.68~bohr$^3$ according to neXDM, a much more reasonable value in
view of the isolated polarizability of \ce{Li+} (0.1930~bohr$^3$, see
Table~\ref{t:ions}). This discrepancy translates directly into the
dispersion coefficients: the XDM $C_6$ for the Li-Li pair in
methyllithium is \SI{127.99}{a.u.} for only \SI{4.02}{a.u.} according
to neXDM, a factor of about 30. The C-C $C_6$ coefficient is also much
lower in XDM (\SI{33.05}{a.u.}) compared to neXDM (\SI{84.05}{a.u.}),
in line with the low charge transfer predicted by the Hirshfeld
partition.

\subsection{Molecular Crystals}
\label{ss:molecular}

\begin{table}
\caption{Mean absolute errors (in kcal/mol per molecule) in the
  lattice energies of the molecular crystals from the X23
  set,\cite{oterodelaroza2012,reilly2013,dolgonos2019} for XDM and
  neXDM with various combinations of method, functional, and basis
  set. Reference values are from Ref.~\citenum{dolgonos2019}.}
\label{t:x23}
\begin{tabular}{llrr}
\hline\hline
 & & XDM & neXDM \\
\hline
\multicolumn{4}{c}{FHI-aims, GGA functionals$^a$} \\
\hline
B86bPBE    & tight        & 0.715 & 0.913 \\
B86bPBE    & lightdense   & 0.825 & 0.700 \\
PBE        & tight        & 1.043 & 0.758 \\
PBE        & lightdense   & 1.135 & 0.828 \\
\hline
\multicolumn{4}{c}{FHI-aims, composites$^{a,b}$} \\
\hline
B86bPBE    & $+$25\% EXX  & 0.657 & 0.679 \\
B86bPBE    & $+$50\% EXX  & 0.695 & 0.716 \\
B86bPBE    & tight basis  & 0.710 & 0.897 \\
PBE        & $+$25\% EXX  & 1.013 & 0.611 \\
PBE        & $+$50\% EXX  & 1.002 & 0.563 \\
PBE        & tight basis  & 1.050 & 0.753 \\
\hline
\multicolumn{4}{c}{Quantum ESPRESSO$^c$} \\
\hline
B86bPBE    & plane waves  & 0.840 & 0.707 \\
PW86PBE    & plane waves  & 0.717 & 0.761 \\
BLYP       & plane waves  & 0.946 & 1.146 \\
PBE        & plane waves  & 1.159 & 0.831 \\
\hline\hline
\end{tabular}
\acnote{$^a$ XDM values from Ref.~\citenum{price2023}.}
\acnote{$^b$ Single point calculations using the method in the second
  column (a hybrid functional or a GGA with a tight basis set)
  calculated at the equilibrium geometry using the GGA in the first
  column plus a lightdense basis set.}
\acnote{$^c$ XDM values calculated in this work.}
\end{table}

We now move on to examine the performance of XDM and neXDM for
molecular crystals. We use as reference the recent paper by Price et
al.\cite{price2023} showing how XDM achieves unprecedented accuracy in
the calculation of lattice energies of molecular crystals, and
recalculate the benchmark sets in that article
(X23\cite{oterodelaroza2012,reilly2013,dolgonos2019},
ICE13\cite{brandenburg2015,dellapia2022}) using neXDM.
Table~\ref{t:x23} shows the mean absolute errors in the lattice
energies of the 23 molecular crystals of the X23 set using XDM and
neXDM. In this case, contrary to previous sections, the crystals and
molecules that enter the calculation have undergone full relaxation at
the same level of theory. XDM and neXDM achieve MAEs that range
between 0.66 and 1.16~kcal/mol for XDM and between 0.56 and 1.15 for
neXDM. The neXDM method significantly improves the performance of the
PBE functional: 1.043 to 0.758~kcal/mol with the tight tier, with
similar improvements for the other GGA functionals and composites.
With B86bPBE, the functional typically paired with XDM, the MAE
decreases with the lightdense basis set (0.825~kcal/mol to
0.700~kcal/mol) and with plane waves (0.840~kcal/mol to
0.707~kcal/mol) but the MAE increases with a tight basis set
(0.715~kcal/mol to 0.913~kcal/mol). To our knowledge, the
B86bPBE-neXDM/lightdense result (or the very similar B86bPBE-neXDM/QE)
is the lowest MAE reported for the X23 set in the literature using a
GGA functional. PW86PBE/QE shows a slight increase in the MAE and BLYP
an increase of 0.2~kcal/mol. Generally, neXDM acts on the X23 data as
an almost uniform increase in binding, and therefore whether it helps
is decided by the underlying functional.

The composite methods, in which a single point with an expensive
method (either a hybrid functional or a larger basis set) is
calculated at the GGA geometry, perform generally better than pure
GGAs. Once more, neXDM greatly improves the performance of
PBE and, in fact, the PBE-50\%-neXDM result at the PBE-neXDM geometry
is the best-performing composite method at 0.563~kcal/mol MAE. In
contrast, the B86bPBE-based composite methods with neXDM show a
slightly higher MAE compared to XDM. However, it is important to note
that, except with regards to the improvement of the PBE-based
functionals, the differences between neXDM and XDM on this set are
likely of the same size as the uncertainty in the benchmark data.

\begin{table}
\caption{Mean absolute errors (in kcal/mol per molecule) in the
  absolute and relative lattice energies of the ice phases from the
  ICE13 set,\cite{brandenburg2015,dellapia2022} for XDM and
  neXDM with various combinations of method, functional, and basis
  set. Reference values from Ref.~\citenum{dellapia2022}.}
\label{t:ice13}
\begin{tabular}{llrrrr}
\hline\hline
 & & \multicolumn{2}{c}{absolute} & \multicolumn{2}{c}{relative} \\
\cline{3-4}\cline{5-6}
 & & XDM & neXDM & XDM & neXDM \\
\hline
\multicolumn{6}{c}{FHI-aims, GGA functionals} \\
\hline
B86bPBE    & tight          & 1.643 & 1.808 & 0.437 & 0.463 \\
B86bPBE    & lightdense     & 2.556 & 2.512 & 0.656 & 0.717 \\
PBE        & tight          & 1.579 & 1.495 & 0.705 & 0.776 \\
PBE        & lightdense     & 2.661 & 2.429 & 0.886 & 0.982 \\
\hline
\multicolumn{6}{c}{FHI-aims, composites$^{a}$} \\
\hline
B86bPBE    & $+$25\% EXX    & 0.797 & 1.256 & 0.473 & 0.395 \\
B86bPBE    & $+$50\% EXX    & 0.427 & 0.153 & 0.303 & 0.230 \\
B86bPBE    & tight basis    & 1.639 & 1.810 & 0.432 & 0.454 \\
PBE        & $+$25\% EXX    & 0.979 & 0.975 & 0.597 & 0.623 \\
PBE        & $+$50\% EXX    & 0.322 & 0.249 & 0.335 & 0.315 \\
PBE        & tight basis    & 1.573 & 1.491 & 0.701 & 0.771 \\
\hline
\multicolumn{6}{c}{Quantum ESPRESSO} \\
\hline
B86bPBE    & plane waves    & 1.545 & 1.768 & 0.547 & 0.527 \\
PW86PBE    & plane waves    & 1.606 & 1.835 & 0.387 & 0.364 \\
BLYP       & plane waves    & 1.949 & 1.975 & 0.240 & 0.286 \\
PBE        & plane waves    & 1.483 & 1.628 & 0.794 & 0.796 \\
\hline\hline
\end{tabular}
\acnote{$^a$ Single point calculations using the method in the second
  column (a hybrid functional or a GGA with a tight basis set)
  calculated at the equilibrium geometry using the GGA in the first
  column plus a lightdense basis set.}
\end{table}

Table~\ref{t:ice13} shows the neXDM and XDM results for the ICE13 set
of 13 ice phases,\cite{brandenburg2015} whose absolute lattice
energies are compared to the diffusion Monte Carlo (DMC) reference
values of Della Pia \emph{et al.}\cite{dellapia2022} The table also
gives the errors in the relative lattice energies, defined as the
differences between all 78 pairs of phases. All GGA functionals
overbind every phase in ICE13, a clear indication of the impact of
delocalization error in these systems.\cite{price2023} Since, as
mentioned for X23, neXDM generally binds molecular
crystals more strongly than XDM, the absolute MAEs tend to increase for neXDM,
except in the case of PBE with the FHI-aims basis sets and
B86bPBE/lightdense. Although neXDM also tends to give slightly worse
MAE for the relative lattice energies, they are at most 0.10~kcal/mol
from their XDM counterparts.

Table~\ref{t:ice13} also shows the results for the composite
functionals, which tend to work better for ice phases because the
inclusion of exact exchange, particularly in functionals with
50\%\ exact exchange, mitigates the effect of delocalization
error. The use of the hybrid functional single point energy removes
the overbinding of the pure GGA functionals, and the two rows with
50\,\% exact exchange show by a wide margin the lowest MAE of the
table, with neXDM improving on XDM in both cases. The
B86bPBE-50\%-neXDM//B86bPBE-neXDM value of 0.153~kcal/mol is, in fact,
the lowest reported MAE for the lattice energies in ICE13 from any
functional,\cite{dellapia2022,price2023} to our knowledge. For the
other entries in the composites part of the table, neXDM causes an
improvement in the absolute lattice energies for the PBE-based
functionals and a deterioration in performance for the B86bPBE-based
functionals, with minor changes in MAE for the relative lattice
energies.

\subsection{Alkali Halides}
\label{ss:alkalihalides}

\begin{figure*}
\includegraphics[width=\textwidth]{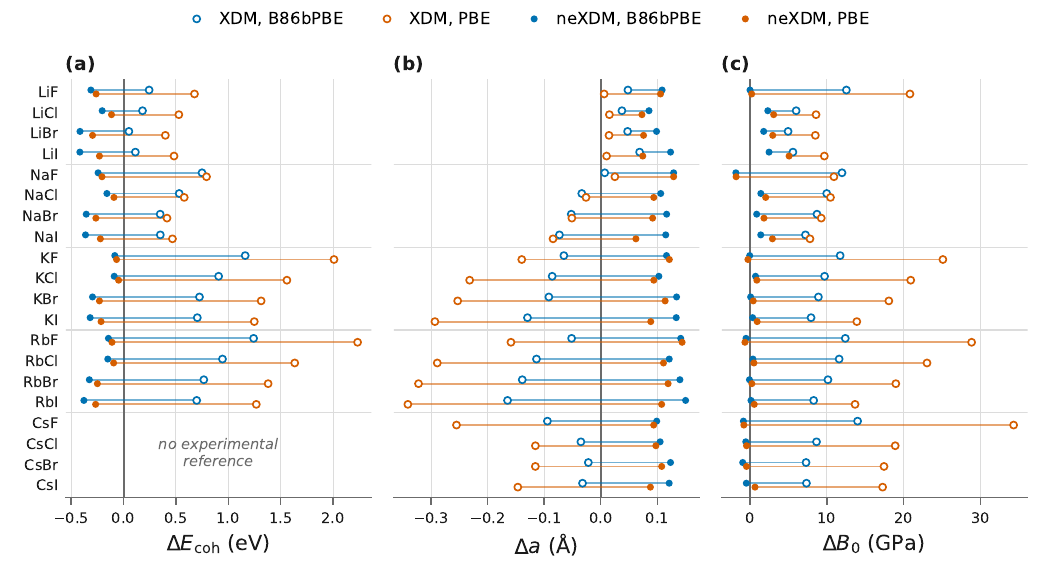}
\caption{Errors with respect to experiment in the (a) cohesive
  energy\cite{doll1997,doll1998}, (b) static lattice
  constants\cite{tao2017,oterodelaroza2020}, and (c) bulk
  moduli\cite{sirdeshmukh2001} of the twenty alkali halides in their
  ambient-conditions phase (B2 for \ce{CsCl}, \ce{CsBr} and \ce{CsI},
  B1 for the rest), calculated with XDM (open circles) and neXDM
  (filled circles) using Quantum ESPRESSO.}
\label{f:alkali}
\end{figure*}

\begin{table}
\caption{Mean absolute errors (MAE) in the cohesive
  energies\cite{doll1997,doll1998}, static lattice
  constants\cite{tao2017,oterodelaroza2020} and bulk
  moduli\cite{sirdeshmukh2001} of the twenty alkali halides
  (ambient-conditions phase$^a$) with XDM and neXDM using Quantum
  ESPRESSO, against experimental data.}
\label{t:alkali}
\begin{tabular}{l@{\hspace{0.6cm}}r@{\hspace{0.6cm}}r}
\hline\hline
 & XDM & neXDM \\
\hline
\multicolumn{3}{c}{B86bPBE} \\
\hline
E$_{\rm coh}$ (eV)     & 0.607 & 0.266 \\
$a$ (\AA)              & 0.070 & 0.118 \\
$B_0$ (GPa)            & 9.25 & 0.86 \\
\hline
\multicolumn{3}{c}{PBE} \\
\hline
E$_{\rm coh}$ (eV)     & 1.061 & 0.186 \\
$a$ (\AA)              & 0.145 & 0.100 \\
$B_0$ (GPa)            & 16.82 & 1.33 \\
\hline\hline
\end{tabular}
\acnote{$^a$ The B2 structure for CsCl, CsBr, and CsI; B1 for every other
  system.}
\end{table}

We now consider the alkali halides, the simplest models of ionic
crystals. In these systems, atoms have approximately unit charge and,
although they are not traditionally associated with dispersion
corrections, their calculated properties are affected by the inclusion
of dispersion effects.\cite{oterodelaroza2020} Figure~\ref{f:alkali}
and Table~\ref{t:alkali} collect the mean absolute errors in the
cohesive energies, lattice constants, and bulk moduli of the twenty
alkali halides, each in its ambient-conditions phase, calculated with
XDM and neXDM using plane waves and PAW pseudopotentials. The values
for the individual salts are given in the supplementary material.  The
cohesive energies are calculated with respect to the neutral free
atoms, in eV per formula unit. The lattice constant and the bulk
modulus both come from the same third-order Birch--Murnaghan fit to a
seven-point $E(V)$ curve centered on the relaxed volume using the
gibbs2 program.\cite{gibbs2a,gibbs2b} It is important to point out
that, unlike TS and similar methods,\cite{bucko2013} the XDM (and
neXDM) forces and stresses are calculated in a manner that is not
completely consistent with the energy. This means that the equilibrium
lattice constants obtained by direct minimization of forces and stress
do not coincide with those obtained from the minimum of the $E(V)$
curve. The reason is that the XDM dispersion coefficients are assumed
to be constant when the corresponding derivatives are taken, which is
not correct. This mismatch results in a disagreement that, for compact
systems such as the alkali halides, can climb to 7\%\ of the
lattice constants.

Regarding the static lattice constants, XDM is about correct for the
four lithium salts and \ce{NaF}, and too short for the other fifteen,
with mean errors of $-0.049$~\AA{} (B86bPBE) and $-0.138$~\AA{} (PBE)
and a worst case of \ce{RbI}, short by 0.165 and 0.341~\AA{}
respectively. neXDM expands every one of the twenty lattices and
overestimates all of them, by $+0.118$~\AA{} (B86bPBE) and
$+0.100$~\AA{} (PBE) on average and by at most 0.150~\AA{}. The two
models are thus comparable in accuracy but opposite in sign, and which
of them is better depends on the functional: the MAE goes from 0.070
to 0.118~\AA{} with B86bPBE and from 0.145 to 0.100~\AA{} with PBE.

In contrast, the improvement of neXDM relative to XDM in the
calculation of bulk moduli is dramatic. XDM roughly doubles the
experimental bulk modulus of the heavy salts and the MAE over the
twenty solids is 9.2~GPa with B86bPBE and 16.8 with PBE. With neXDM
these MAE fall to 0.86 and 1.33~GPa, a factor of eleven and thirteen,
with no error larger than 2.5~GPa (B86bPBE) or 5.1~GPa (PBE). The
results for the cohesive energies show a similar pattern. XDM
overbinds every salt by an amount that grows with the size of the
cation: with B86bPBE the error is between 0.05 and 0.24~eV for the
lithium salts, 0.35 and 0.75~eV for the sodium salts, and 0.70 and
1.24~eV for the potassium and rubidium ones, and with PBE it reaches
2.23~eV for \ce{RbF}. In contrast, with neXDM the MAEs over the
sixteen salts fall from 0.607 to 0.266~eV (B86bPBE) and from 1.061 to
0.186~eV (PBE), with slight underbinding.

The improved performance of neXDM relative to XDM mirrors almost
exactly that of TS/HI over TS, demonstrated by Bu\v{c}ko et
al.\cite{bucko2013b}, and it can be traced back to the ion
polarizabilities. Summed over one formula unit, the XDM
polarizabilities are between 148 and 406~bohr$^3$, whereas neXDM gives
between 15 and 71~bohr$^3$, in good agreement with the sum of the
free-ion polarizabilities of Gould and Bu\v{c}ko\cite{gould2016}
(15--77~bohr$^3$). The XDM values are not simply too large, they are
also unphysical. The Clausius--Mossotti relation:
\begin{equation}
\label{eq:clausius}
\frac{\epsilon_\infty - 1}{\epsilon_\infty + 2} = \frac{4\pi \sum_i \alpha_i}{3V}
\end{equation}
has no solution for any of the twenty alkali halides because
the right-hand side exceeds unity, resulting in a polarization
catastrophe and the prediction of a ferroelectric solid.

\begin{table}
\caption{Errors in the static polarizability of 16 alkali halide salts
  (Li--Rb, F--I, in bohr$^3$) for XDM and neXDM and for the dispersion
  models benchmarked by Caldeweyher \emph{et
  al.}\cite{caldeweyher2020} against the experimentally derived values
  also reported in their work.}
\label{t:saltalpha}
\small
\setlength{\tabcolsep}{2pt}
\begin{tabular}{lrr}
\hline\hline
method & MD & MAD \\
       &    &     \\
\hline

D4\cite{caldeweyher2019,caldeweyher2020} & -0.2 & 1.7 \\
neXDM, B86bPBE      & 11.1 & 11.1 \\
neXDM, PBE          & 11.2 & 11.2 \\
MBD/FI\cite{gould2016}       & 14.6 & 14.6 \\
D3\cite{grimme2010,grimme2011b} & 31.8 & 31.8 \\
MBD/HI\cite{bucko2014,gould2016} & 32.0 & 32.0 \\
MBD/HI\cite{bucko2014,gould2016} & 32.9 & 32.9 \\
TS/HI\cite{bucko2013b,bucko2014} & 33.7 & 33.7 \\
MBD\cite{tkatchenko2012,ambrosetti2014}  & 164.9 & 165.6 \\
TS\cite{tkatchenko2009}      & 192.0 & 192.0 \\
XDM, B86bPBE        & 200.9 & 200.9 \\
XDM, PBE            & 202.1 & 202.1 \\
dDsC\cite{steinmann2011}     & 256.6 & 256.6 \\
\hline\hline
\end{tabular}
\end{table}

To illustrate this point, Table~\ref{t:saltalpha} shows the average
errors in the calculation of the static polarizabilities of the
sixteen alkali halides studied by Caldeweyher \emph{et
al.}\cite{caldeweyher2020} In their work, they compared a range of
dispersion models against the polarizabilities derived from the
refractive index through the Clausius--Mossotti relation
(Eq.~\ref{eq:clausius}).\cite{tessman1953} neXDM gives a mean absolute
deviation of 11.1~bohr$^3$ (B86bPBE) and 11.2~bohr$^3$ (PBE), second
only to D4 (1.7~bohr$^3$) and ahead of MBD/FI (14.6), D3 (31.8),
MBD/HI (32.0), MBD/HI (32.9), and TS/HI (33.7). XDM gives 200.9 and
202.1~bohr$^3$ and ranks as one of the worst-performing functionals
besides MBD (165.6), TS (192.0), and dDsC (256.6).

\subsection{Layered Materials}

\begin{table*}
\caption{Errors in the exfoliation energies ($E_{\rm exfol}$) and
  interlayer spacings ($d$) of the 26 layered
  solids.\cite{bjorkman2014,hummel2016} The statistics with neXDM and
  other methods, and various combinations of functional and basis set
  are shown, compared to the indicated values from the literature.
}
\label{t:layer}
\scriptsize
\begin{tabular}{lll@{\hspace{0.6cm}}rr@{\hspace{0.6cm}}rr}
\hline\hline
 & & & \multicolumn{2}{c}{$E_{\rm exfol}$ (meV/\AA$^2$)}
 & \multicolumn{2}{c}{$d$ (\AA)} \\
\cline{4-5}\cline{6-7}
Model & Functional & Basis & \multicolumn{1}{c}{ME} & \multicolumn{1}{c}{MAE} & \multicolumn{1}{c}{ME} & \multicolumn{1}{c}{MAE} \\
\hline
\multicolumn{7}{c}{This work} \\
\hline
neXDM           & B86bPBE    & lightdense   &  $-1.27$ &  2.36 & $+0.123$ & 0.139 \\
neXDM           & B86bPBE    & tight        &  $-1.25$ &  2.25 & $+0.092$ & 0.125 \\
neXDM           & PBE        & lightdense   &  $-0.17$ &  2.72 & $+0.108$ & 0.138 \\
neXDM           & PBE        & tight        &  $-1.20$ &  2.49 & $+0.084$ & 0.121 \\
XDM                      & B86bPBE    & lightdense   &   4.82 &  4.83 & $-0.048$ & 0.080 \\
XDM                      & B86bPBE    & tight        &   7.13 &  7.13 & $-0.114$ & 0.116 \\
XDM                      & PBE        & lightdense   &   4.69 &  4.86 & $-0.036$ & 0.073 \\
XDM                      & PBE        & tight        &   5.53 &  5.69 & $-0.081$ & 0.093 \\
MBD-NL                   & PBE        & lightdense   &  $-4.63$ &  4.63 & $+0.079$ & 0.123 \\
MBD-NL                   & PBE        & tight        &  $-5.40$ &  5.40 & $+0.067$ & 0.113 \\
MBD-NL$^{\rm surf}$      & PBE        & lightdense   &  $-5.06$ &  5.06 & $+0.091$ & 0.127 \\
MBD-NL$^{\rm surf}$      & PBE        & tight        &  $-5.81$ &  5.81 & $+0.080$ & 0.124 \\
TS                       & PBE        & lightdense   &  13.13 & 13.13 & $-0.113$ & 0.159 \\
TS                       & PBE        & tight        &  12.71 & 12.71 & $-0.142$ & 0.170 \\
\hline
\multicolumn{7}{c}{Rumson et al.\ (Ref.~\citenum{rumson2026})} \\
\hline
XDM(BJ)                  & B86bPBE    & lightdenser  &   5.88 &  5.89 & $-0.05$ & 0.07 \\
XDM(BJ)+ATM              & B86bPBE    & lightdenser  &   1.18 &  2.30 & $-0.01$ & 0.07 \\
XDM(BJ)                  & B86bPBE    & tight        &   7.67 &  7.67 & $-0.12$ & 0.12 \\
XDM(BJ)+ATM              & B86bPBE    & tight        &   3.17 &  3.49 & $-0.07$ & 0.08 \\
XDM(BJ)                  & B86bPBE    & PW           &   3.30 &  4.67 & $-0.04$ & 0.09 \\
XDM(BJ)+ATM              & B86bPBE    & PW           &  $-0.18$ &  3.05 & $-0.01$ & 0.08 \\
XDM(BJ)                  & PBE        & lightdenser  &   5.47 &  5.63 & $-0.03$ & 0.06 \\
XDM(BJ)+ATM              & PBE        & lightdenser  &   0.87 &  2.67 & $+0.01$ & 0.07 \\
XDM(BJ)                  & PBE        & tight        &   6.71 &  6.84 & $-0.09$ & 0.11 \\
XDM(BJ)+ATM              & PBE        & tight        &   2.12 &  3.33 & $-0.05$ & 0.09 \\
XDM(BJ)                  & PBE        & PW           &   1.65 &  4.39 & $+0.05$ & 0.13 \\
XDM(BJ)+ATM              & PBE        & PW           &  $-2.21$ &  3.86 & $+0.11$ & 0.15 \\
\hline\hline
\end{tabular}
\end{table*}

Lastly, we examine the performance of neXDM in layered materials:
\ce{h-BN}, graphite, PbO, and transition metal dichalcogenides. In
these systems, binding arises almost entirely from dispersion
interactions.  Table~\ref{t:layer} shows the errors in the exfoliation
energies and interlayer spacings of the 26 layered solids of the LM26
set against the random-phase-approximation (RPA) reference values of
Bj\"orkman,\cite{bjorkman2014} the only exception being \ce{h-BN}, for
which the second-order M{\o}ller--Plesset value of Hummel \emph{et
al.}\cite{hummel2016} is used.\cite{oterodelaroza2020b} The
exfoliation energy is obtained from the minimum of an
interlayer-distance scan at fixed in-plane lattice constants. We only
considered GGA functionals in this case, as we experienced severe
difficulties with the SCF convergence and the stabilization of the
correct magnetic state using hybrid functionals. Some other dispersion
corrections available in the FHI-aims code have been calculated as
well using the default settings. PW denotes plane waves with PAW
pseudopotentials (Quantum ESPRESSO), and the ATM rows include the
three-body Axilrod--Teller--Muto term.

Regarding the interlayer spacings, the two models behave as in the
alkali halides (Sec.~\ref{ss:alkalihalides}), with errors of similar
size and opposite sign. XDM underestimates the separation, with mean
errors between $-0.04$ and $-0.11$~\AA{}, and neXDM overestimates it
by $+0.08$ to $+0.12$~\AA{}. The MAEs are 0.073--0.116~\AA{} for XDM
and 0.121--0.139~\AA{} for neXDM. Both are well within the 0.2~\AA{}
that is usually considered acceptable for these
systems.\cite{oterodelaroza2020b}

In agreement with previous reports in the
literature,\cite{oterodelaroza2020b} XDM overbinds the exfoliation
energies in this set (Table~\ref{t:layer}). The mean error from XDM is
between $+4.7$ and $+7.1$~meV/\AA$^2$ on exfoliation energies whose
reference values go from 16 to 40~meV/\AA$^2$, a mean relative error
of 25 to 37\,\%. These errors cannot be explained from underlying
deficiencies in the base functional\cite{oterodelaroza2020b}: coupled
with the same functional (PBE), XDM and TS overbind while MBD-NL
underbinds.

The use of the neXDM model greatly reduces errors in the calculated
exfoliation energies. The mean errors of the four neXDM rows lie
between $-0.17$ and $-1.27$~meV/\AA$^2$ and the mean absolute errors
fall to 2.25~meV/\AA$^2$ (B86bPBE/tight), 2.36 (B86bPBE/lightdense),
2.49 (PBE/tight) and 2.72 (PBE/lightdense), against 4.83 to 7.13 for
XDM.  An analysis of the data shows that the same mechanism is at work
here as in the ionic systems of the preceding sections. Of the 26
materials, XDM works well only on graphite and \ce{h-BN}: with PBE and
the tight species defaults their mean absolute error is
1.0~meV/\AA$^2$ (4.9\,\% in relative terms). For the other 23
transition-metal dichalcogenides and PbO, in which a metal cation is paired
with a polarizable chalcogen anion, XDM gives a mean absolute
error of 6.1~meV/\AA$^2$ (32.0\,\%) and neXDM reduces it to
2.5~meV/\AA$^2$ (12.4\,\%). The extreme case is \ce{ZrTe2}, overbound
by 13.6~meV/\AA$^2$ with XDM and by 5.9 with neXDM. The pattern
repeats through the whole series: every selenide and telluride of
\ce{Ti}, \ce{Zr}, \ce{Hf}, \ce{Nb} and \ce{Ta} is overbound by between
5.6 and 13.6~meV/\AA$^2$ with XDM, and none of them is in error by
more than 5.9 with neXDM.

It is worth comparing these numbers with the recent work of Rumson
\emph{et al.},\cite{rumson2026} who addressed the same overbinding of
XDM relative to the RPA by adding the three-body
Axilrod--Teller--Muto (ATM) term. Without the ATM term their XDM(BJ) results
reproduce the overbinding described above, with mean absolute errors
of 4.4 to 7.7~meV/\AA$^2$ for PBE and B86bPBE across three basis
settings. The ATM term, which is repulsive
for the nearly equilateral atom triples that span adjacent layers,
lowers the mean absolute errors to 3.1--3.5 with plane waves and the
tight species defaults and to 2.3--2.7 with their lightdenser
settings, the latter helped by basis-set superposition error according
to the authors. At matched basis sets neXDM is as accurate or better:
2.25 (B86bPBE) and 2.49~meV/\AA$^2$ (PBE) with the tight defaults,
against 3.5 and 3.3 for XDM(BJ)+ATM.

The use of the ATM contribution in XDM was considered in our previous
work.\cite{xdmc9} The difficulty in its application, which we think is
also present when ATM is used in combination with other pairwise
dispersion corrections (e.g.\ D3), arises from the fact that, while
the ATM contribution is a physical effect and the asymptotic
expression is known, it is unclear how to damp it in the overlapping
regime. In other words, the size of the ATM term relative to the
pairwise contribution depends crucially on how it is damped, a choice
that is ultimately arbitrary. Since the inclusion of the ATM term
almost always results in increased repulsion, there is a temptation to
use its absence as a blanket justification in those cases when a
dispersion correction overbinds, for whatever reason.

\begin{table}
\caption{Polarizability per formula unit (in bohr$^3$) of
  the eight semiconducting transition-metal dichalcogenides of the LM26
  set. The CM entries were obtained from the in-plane and out-of-plane
  electronic dielectric constants of the bulk crystals calculated by
  Laturia \emph{et al.}\cite{laturia2018} through the
  Clausius--Mossotti relation (Eq.~\ref{eq:clausius}), averaging the
  three Cartesian components, at the cell volume of that work. The XDM
  and neXDM columns are the sum of the atomic polarizabilities over the
  formula unit (PBE, lightdense, reference interlayer spacing).}
\label{t:layercm}
\begin{tabular}{lrrr}
\hline\hline
 & CM & XDM & neXDM \\
\hline
\ce{HfS2}  &  75.5 & 144.9 &  75.8 \\
\ce{HfSe2} &  93.7 & 157.5 &  71.4 \\
\ce{ZrS2}  &  79.5 & 155.1 &  81.3 \\
\ce{MoS2}  &  67.5 & 126.8 &  76.1 \\
\ce{MoSe2} &  80.8 & 138.7 &  70.1 \\
\ce{MoTe2} & 105.3 & 163.3 &  97.5 \\
\ce{WS2}   &  66.6 & 113.4 &  74.9 \\
\ce{WSe2}  &  78.4 & 125.7 &  71.9 \\
\hline
ME (bohr$^3$) & & 59.8 & $-3.5$ \\
MAE (bohr$^3$) & & 59.8 & 8.3 \\
MARE (\%) & & 75.0 & 10.1 \\
\hline\hline
\end{tabular}
\end{table}

Table~I in Ref.~\citenum{rumson2026} shows a degradation in the XDM
parametrization statistics when the ATM term is included, and the
preceding results for the alkali halides in combination with the
excellent performance of neXDM for the exfoliation energies suggest
the XDM errors in the LM26 set also originate from the neutral-atom
reference in the Hirshfeld partition and the linear volume scaling of
the atom-in-molecule polarizabilities. To illustrate this point,
Table~\ref{t:layercm} compares the polarizability per formula unit
derived from the dielectric constants of Laturia \emph{et
al.}\cite{laturia2018} through the Clausius--Mossotti relation
(Eq.~\ref{eq:clausius}) with the sum of the atomic polarizabilities of
XDM and neXDM for the eight semiconducting dichalcogenides in the set,
for which such dielectric constants are available. The neXDM atomic
polarizabilities agree with the Clausius--Mossotti value with an
MAE of 8.3~a.u. and an MARE of 10.1\%. Similar to its behavior for the
alkali halides, XDM overestimates the polarizabilities by a factor
between 55\,\% (\ce{MoTe2}) and 95\,\% (\ce{ZrS2}), with an MAE of
59.8~a.u. and a 75\%\ MARE.

\section{Conclusions}
\label{s:conclusions}

In this article, we proposed a new dispersion model built on the
modification of the exchange-hole dipole moment (XDM) model called
non-empirical XDM (neXDM). We modified XDM in two ways. First, neXDM
replaces the Hirshfeld partitioning scheme with its iterative
counterpart, to make the partitioning weights aware of the charge of
the system and prevent the bias towards the neutral atom
reference. Second, the calculation of atom-in-molecule
polarizabilities no longer uses the experimental static
polarizabilities and the linear atomic volume scaling. Instead, we use
the Kirkwood formula to calculate the atom-in-molecule
polarizabilities. With these two changes, the dispersion energy is
determined entirely by the self-consistent density, its derivatives,
and the kinetic energy density, making neXDM a pure meta-GGA
dispersion functional. The method was implemented in the postg,
FHI-aims, and Quantum ESPRESSO programs, and the damping parameters
were refitted for a collection of base functionals.

For neutral molecules and molecular crystals, neXDM performs on par
with XDM. The molecular $C_6$ coefficients of the 48 species with
dipole oscillator strength distributions are reproduced with a mean
absolute relative error of 10.3\,\%, against 10.5\,\% for XDM. The
KB49 damping-parameter fits give root-mean-square percent deviations
that differ from those of XDM by a few tenths of a percent for almost
every combination of functional, code, and basis set, and the average
MAEs over sixteen functionals on the S22x5, S66x8, IHB100x10, and
SH250x10 sets differ from XDM by a few hundredths of a kcal/mol, which
is less than the spread between functionals. On the GMTKN55 database,
the average WTMAD-2 is 8.46~kcal/mol with neXDM and 8.51 with XDM. On
the X23 set of molecular crystals, neXDM with B86bPBE and the
lightdense species defaults achieves an MAE of 0.700~kcal/mol, the
lowest reported for this set with a GGA functional, and the composite
PBE-50X-neXDM//PBE-neXDM method gives 0.563~kcal/mol. On the ICE13
set, the B86bPBE-50X-neXDM//B86bPBE-neXDM composite gives an MAE of
0.153~kcal/mol in the absolute lattice energies, the lowest reported
for any functional, although for the pure GGAs neXDM binds the ice
phases slightly more strongly than XDM and the absolute MAEs increase
by up to 0.23~kcal/mol.

For systems with charged atoms, the neXDM improvement over XDM is
dramatic. The static polarizabilities and $C_6$ coefficients of the
alkali cations and halide anions, which XDM overestimates by up to an
order of magnitude, are reproduced by neXDM with mean absolute
relative errors of 39 and 34\% instead of 500 and 449\%. This
translates directly into the energetics: the MAE of the IONPI19 set
falls from 1.772 to 1.473~kcal/mol, that of its alkali cation--$\pi$
subset from 1.728 to 0.929~kcal/mol, and that of the ALK8 subset of
GMTKN55 from 16.01 to 6.22~kcal/mol. In the twenty alkali halides, XDM
predicts lattice constants that are too short, bulk moduli that are
roughly twice the experimental values, and cohesive energies that are
too large by up to 2.2~eV, and its formula-unit polarizabilities are
so large that the Clausius--Mossotti relation has no solution and
every salt is predicted to be ferroelectric. neXDM corrects all of
these problems: the MAEs in the bulk moduli fall from 9.25 to 0.86~GPa
(B86bPBE) and from 16.82 to 1.33~GPa (PBE), those in the cohesive
energies from 0.607 to 0.266~eV and from 1.061 to 0.186~eV, and the
mean absolute deviation of the salt polarizabilities from the
experimentally derived values falls from 201 to 11~bohr$^3$, second
only to D4 among the dispersion models reported.

The layered materials follow the same pattern. XDM overbinds the 26
solids of the LM26 set by 25 to 37\,\% relative to the RPA reference
values.  With neXDM the MAE in the exfoliation energies falls from
4.8--7.1 to 2.25--2.72~meV/\AA$^2$, throwing into question the
putative need of a three-body dispersion term to correctly model these
systems. We show that the origin of the improvement is the same as in
the alkali halides: the XDM polarizabilities of the metal atoms are
those of the neutral free atoms, whereas the neXDM values are
consistent with the free-ion polarizabilities at the iterative
Hirshfeld charge, and the sum of the neXDM atomic polarizabilities
over the formula unit reproduces the electronic polarizability of the
eight semiconducting dichalcogenides derived from their dielectric
constants with an MAE of 8.3~bohr$^3$ (10\,\%), against 59.8~bohr$^3$
(75\,\%) for XDM.

In summary, neXDM removes the experimental free-atom polarizabilities
and the neutral-atom reference from XDM, retains the accuracy of XDM
for neutral molecules and molecular crystals, and greatly improves
upon XDM for ionic molecules, ionic solids, and layered materials with
charged atoms, for which XDM and the other Hirshfeld-based dispersion
corrections fail qualitatively. The Kirkwood formula is not an
accurate polarizability functional, and more research is needed to
improve this method even further. Nevertheless, we believe this is an
important step towards a dispersion correction that is accurate,
non-empirical, physics-based, and applicable to every kind of chemical
system on an equal footing.

\section*{Supplementary Material}

The supplementary material contains: i) the mean absolute errors of
XDM and neXDM on the five non-covalent interaction benchmark sets for
each of the sixteen functionals, ii) the WTMAD-2 values for the five
categories of GMTKN55 and for the database as a whole, iii) the mean
absolute deviations of both models on each of the fifty GMTKN55
subsets examined, and iv) the cohesive energies, static lattice constants
and static bulk moduli of the twenty individual alkali halides. The
supplementary material also contains the damping parameters of every
neXDM parametrization carried out in this work, including the FHI-aims
species defaults not listed in Table~\ref{t:params} and the Quantum
ESPRESSO sets at every combination of the two density options,
together with their KB49 statistics, as well as an archive with the
743 numerical reference atomic densities, the coefficients and
exponents of their Slater-type fits (Eq.~\ref{eq:rhofit}), and the
plots comparing the fitted and the numerical densities.

\begin{acknowledgments}
AOR thanks the Spanish Ministerio de Ciencia, Innovaci\'on y
Universidades and the Agencia Estatal de Investigaci\'on, project
PID2024-158791NB-I00\allowbreak{}
(MICIU/\allowbreak AEI/\allowbreak 10.13039/\allowbreak 501100011033),
and the Principality of Asturias, project IDE/2024/000726. All projects are cofinanced by EU
FEDER funds. The authors are grateful to the Digital Research Alliance
of Canada and to the MALTA Consolider supercomputing center for
computational resources.
\end{acknowledgments}

\section*{Author Declarations}

\subsection*{Conflict of Interest}

The authors have no conflicts to disclose.

\subsection*{Author Contributions}

\textbf{Alastair J. A. Price:} Conceptualization (equal); formal analysis
(equal); investigation (equal); methodology (equal); software (equal);
validation (equal); writing -- review and editing (equal).

\textbf{Alberto Otero-de-la-Roza:} Conceptualization (equal); formal analysis
(equal); investigation (equal); methodology (equal); software (equal);
validation (equal); writing -- original draft (lead); writing --
review and editing (equal).

\section*{Data Availability}

The data that support the findings of this study are available within
the article and its supplementary material.

\cleardoublepage

\bibliographystyle{aipnum4-2}
\bibliography{ixdm}

@article{bucko2014,
  author    = {Bu{\v{c}}ko, Tom{\'a}{\v{s}} and Leb{\`e}gue, S{\'e}bastien and {\'A}ngy{\'a}n, J{\'a}nos G. and Hafner, J{\"u}rgen},
  title     = {Extending the applicability of the {Tkatchenko-Scheffler} dispersion correction via iterative Hirshfeld partitioning},
  journal   = {J. Chem. Phys.},
  volume    = {141},
  pages     = {034114},
  year      = {2014},
  doi       = {10.1063/1.4890003},
  url       = {https://doi.org/10.1063/1.4890003},
}

@article{hirshfeld1977,
  author    = {Hirshfeld, F. L.},
  title     = {Bonded-atom fragments for describing molecular charge densities},
  journal   = {Theor. Chim. Acta},
  volume    = {44},
  pages     = {129--138},
  year      = {1977},
  doi       = {10.1007/BF00549096},
  url       = {https://doi.org/10.1007/BF00549096},
}

@article{bultinck2007,
  author    = {Bultinck, Patrick and Van Alsenoy, Christian and Ayers, Paul W. and Carb{\'o}-Dorca, Ramon},
  title     = {Critical analysis and extension of the Hirshfeld atoms in molecules},
  journal   = {J. Chem. Phys.},
  volume    = {126},
  pages     = {144111},
  year      = {2007},
  doi       = {10.1063/1.2715563},
  url       = {https://doi.org/10.1063/1.2715563},
}

@article{vanpoucke2013,
  author    = {Vanpoucke, Danny E. P. and Bultinck, Patrick and Van Driessche, Isabel},
  title     = {Extending {Hirshfeld-I} to bulk and periodic materials},
  journal   = {J. Comput. Chem.},
  volume    = {34},
  pages     = {405--417},
  year      = {2013},
  doi       = {10.1002/jcc.23088},
  url       = {https://doi.org/10.1002/jcc.23088},
}

@article{tkatchenko2009,
  author    = {Tkatchenko, Alexandre and Scheffler, Matthias},
  title     = {Accurate Molecular Van Der Waals Interactions from {Ground-State} Electron Density and {Free-Atom} Reference Data},
  journal   = {Phys. Rev. Lett.},
  volume    = {102},
  pages     = {073005},
  year      = {2009},
  doi       = {10.1103/PhysRevLett.102.073005},
  url       = {https://doi.org/10.1103/PhysRevLett.102.073005},
}

@article{bucko2013,
  author    = {Bu{\v{c}}ko, Tom{\'a}{\v{s}} and Leb{\`e}gue, S. and Hafner, J{\"u}rgen and {\'A}ngy{\'a}n, J. G.},
  title     = {{Tkatchenko-Scheffler} van der Waals correction method with and without self-consistent screening applied to solids},
  journal   = {Phys. Rev. B},
  volume    = {87},
  pages     = {064110},
  year      = {2013},
  doi       = {10.1103/PhysRevB.87.064110},
  url       = {https://doi.org/10.1103/PhysRevB.87.064110},
}

@article{bucko2013b,
  author    = {Bu{\v{c}}ko, Tom{\'a}{\v{s}} and Leb{\`e}gue, S{\'e}bastien and Hafner, J{\"u}rgen and {\'A}ngy{\'a}n, J{\'a}nos G.},
  title     = {Improved Density Dependent Correction for the Description of London Dispersion Forces},
  journal   = {J. Chem. Theory Comput.},
  volume    = {9},
  pages     = {4293--4299},
  year      = {2013},
  doi       = {10.1021/ct400694h},
  url       = {https://doi.org/10.1021/ct400694h},
}

@article{verstraelen2016,
  author    = {Verstraelen, Toon and Vandenbrande, Steven and Heidar-Zadeh, Farnaz and Vanduyfhuys, Louis and Van Speybroeck, Veronique and Waroquier, Michel and Ayers, Paul W.},
  title     = {Minimal Basis Iterative Stockholder: Atoms in Molecules for {Force-Field} Development},
  journal   = {J. Chem. Theory Comput.},
  volume    = {12},
  pages     = {3894--3912},
  year      = {2016},
  doi       = {10.1021/acs.jctc.6b00456},
  url       = {https://doi.org/10.1021/acs.jctc.6b00456},
}

@article{gould2016b,
  author    = {Gould, T. and Bu{\v c}ko, T.},
  title     = {{C}$_6$ Coefficients and Dipole Polarizabilities for All Atoms and Many Ions in Rows 1--6 of the Periodic Table},
  journal   = {J. Chem. Theory Comput.},
  volume    = {12},
  pages     = {3603--3613},
  year      = {2016},
  doi       = {10.1021/acs.jctc.6b00361},
  url       = {https://doi.org/10.1021/acs.jctc.6b00361},
}

@article{gould2016c,
  author    = {Gould, Tim},
  title     = {How polarizabilities and {C}$_6$ coefficients actually vary with atomic volume},
  journal   = {J. Chem. Phys.},
  volume    = {145},
  pages     = {084308},
  year      = {2016},
  doi       = {10.1063/1.4961643},
  url       = {https://doi.org/10.1063/1.4961643},
}

@article{gould2016,
  author    = {Gould, Tim and Leb{\`e}gue, S{\'e}bastien and {\'A}ngy{\'a}n, J{\'a}nos G. and Bu{\v c}ko, Tom{\'a}{\v s}},
  title     = {A Fractionally Ionic Approach to Polarizability and van der {W}aals Many-Body Dispersion Calculations},
  journal   = {J. Chem. Theory Comput.},
  volume    = {12},
  pages     = {5920--5930},
  year      = {2016},
  doi       = {10.1021/acs.jctc.6b00925},
  url       = {https://doi.org/10.1021/acs.jctc.6b00925},
}

@article{andersen1999,
  author  = {Andersen, T. and Haugen, H. K. and Hotop, H.},
  title   = {Binding Energies in Atomic Negative Ions: {III}},
  journal = {J. Phys. Chem. Ref. Data},
  volume  = {28},
  pages   = {1511--1533},
  year    = {1999},
  doi     = {10.1063/1.556047},
}

@article{rienstrakiracofe2002,
  author  = {Rienstra-Kiracofe, Jonathan C. and Tschumper, Gregory S. and
             Schaefer, Henry F. and Nandi, Sreela and Ellison, G. Barney},
  title   = {Atomic and Molecular Electron Affinities: Photoelectron
             Experiments and Theoretical Computations},
  journal = {Chem. Rev.},
  volume  = {102},
  pages   = {231--282},
  year    = {2002},
  doi     = {10.1021/cr990044u},
}

@misc{kramida2024,
  author       = {Kramida, A. and Ralchenko, Yu. and Reader, J. and {NIST ASD
                  Team}},
  title        = {{NIST} Atomic Spectra Database (version 5.11)},
  howpublished = {National Institute of Standards and Technology,
                  Gaithersburg, MD},
  year         = {2024},
  note         = {\url{https://physics.nist.gov/asd}},
  doi          = {10.18434/T4W30F},
}

@article{b86b,
  author  = {Becke, A. D.},
  title   = {On the large-gradient behavior of the density functional exchange
             energy},
  journal = {J. Chem. Phys.},
  volume  = {85},
  pages   = {7184--7187},
  year    = {1986},
  doi     = {10.1063/1.451353},
}

@article{pbe,
  author  = {Perdew, John P. and Burke, Kieron and Ernzerhof, Matthias},
  title   = {Generalized Gradient Approximation Made Simple},
  journal = {Phys. Rev. Lett.},
  volume  = {77},
  pages   = {3865--3868},
  year    = {1996},
  doi     = {10.1103/PhysRevLett.77.3865},
}

@article{giannozzi2009,
  author  = {Giannozzi, Paolo and Baroni, Stefano and Bonini, Nicola and
             Calandra, Matteo and Car, Roberto and Cavazzoni, Carlo and
             Ceresoli, Davide and Chiarotti, Guido L. and Cococcioni, Matteo
             and Dabo, Ismaila and Dal Corso, Andrea and de Gironcoli, Stefano
             and Fabris, Stefano and Fratesi, Guido and Gebauer, Ralph and
             Gerstmann, Uwe and Gougoussis, Christos and Kokalj, Anton and
             Lazzeri, Michele and Martin-Samos, Layla and Marzari, Nicola and
             Mauri, Francesco and Mazzarello, Riccardo and Paolini, Stefano
             and Pasquarello, Alfredo and Paulatto, Lorenzo and Sbraccia,
             Carlo and Scandolo, Sandro and Sclauzero, Gabriele and Seitsonen,
             Ari P. and Smogunov, Alexander and Umari, Paolo and Wentzcovitch,
             Renata M.},
  title   = {{QUANTUM ESPRESSO}: a modular and open-source software project
             for quantum simulations of materials},
  journal = {J. Phys.: Condens. Matter},
  volume  = {21},
  pages   = {395502},
  year    = {2009},
  doi     = {10.1088/0953-8984/21/39/395502},
}

@article{giannozzi2017,
  author  = {Giannozzi, P. and Andreussi, O. and Brumme, T. and Bunau, O. and
             Buongiorno Nardelli, M. and Calandra, M. and Car, R. and
             Cavazzoni, C. and Ceresoli, D. and Cococcioni, M. and Colonna, N.
             and Carnimeo, I. and Dal Corso, A. and de Gironcoli, S. and
             Delugas, P. and DiStasio, R. A. and Ferretti, A. and Floris, A.
             and Fratesi, G. and Fugallo, G. and Gebauer, R. and Gerstmann, U.
             and Giustino, F. and Gorni, T. and Jia, J. and Kawamura, M. and
             Ko, H.-Y. and Kokalj, A. and K\"u\c{c}\"ukbenli, E. and Lazzeri,
             M. and Marsili, M. and Marzari, N. and Mauri, F. and Nguyen, N.
             L. and Nguyen, H.-V. and Otero-de-la-Roza, A. and Paulatto, L.
             and Ponc\'e, S. and Rocca, D. and Sabatini, R. and Santra, B. and
             Schlipf, M. and Seitsonen, A. P. and Smogunov, A. and Timrov, I.
             and Thonhauser, T. and Umari, P. and Vast, N. and Wu, X. and
             Baroni, S.},
  title   = {Advanced capabilities for materials modelling with {Quantum
             ESPRESSO}},
  journal = {J. Phys.: Condens. Matter},
  volume  = {29},
  pages   = {465901},
  year    = {2017},
  doi     = {10.1088/1361-648X/aa8f79},
}

@article{koelling1977,
  author  = {Koelling, D. D. and Harmon, B. N.},
  title   = {A technique for relativistic spin-polarised calculations},
  journal = {J. Phys. C: Solid State Phys.},
  volume  = {10},
  pages   = {3107--3114},
  year    = {1977},
  doi     = {10.1088/0022-3719/10/16/019},
}

@article{clementi1974,
  author  = {Clementi, E. and Roetti, C.},
  title   = {Roothaan-{Hartree}-{Fock} atomic wavefunctions: Basis functions
             and their coefficients for ground and certain excited states of
             neutral and ionized atoms, {$Z \leq 54$}},
  journal = {At. Data Nucl. Data Tables},
  volume  = {14},
  pages   = {177--478},
  year    = {1974},
  doi     = {10.1016/S0092-640X(74)80016-1},
}

@article{koga1997b,
  author  = {Koga, Toshikatsu and Matsuyama, Hisashi},
  title   = {Analytical {Hartree-Fock} electron densities for singly charged
             cations and anions},
  journal = {Theor. Chem. Acc.},
  year    = {1997},
  volume  = {98},
  pages   = {129--136},
  doi     = {10.1007/s002140050286},
}

@article{koga1999,
  author  = {Koga, Toshikatsu and Kanayama, Katsutoshi and Watanabe, Takahiro
             and Thakkar, Ajit J.},
  title   = {Analytical {Hartree-Fock} wave functions subject to cusp and
             asymptotic constraints: {He} to {Xe}, {Li$^+$} to {Cs$^+$},
             {H$^-$} to {I$^-$}},
  journal = {Int. J. Quantum Chem.},
  volume  = {71},
  pages   = {491--497},
  year    = {1999},
  doi     = {10.1002/(SICI)1097-461X(1999)71:6<491::AID-QUA6>3.0.CO;2-T},
}

@article{baerends1973,
  author  = {Baerends, E. J. and Ellis, D. E. and Ros, P.},
  title   = {Self-consistent molecular {Hartree-Fock-Slater} calculations
             {I}. The computational procedure},
  journal = {Chem. Phys.},
  volume  = {2},
  pages   = {41--51},
  year    = {1973},
  doi     = {10.1016/0301-0104(73)80059-X},
}

@article{dunlap1979,
  author  = {Dunlap, B. I. and Connolly, J. W. D. and Sabin, J. R.},
  title   = {On first-row diatomic molecules and local density models},
  journal = {J. Chem. Phys.},
  volume  = {71},
  pages   = {4993--4999},
  year    = {1979},
  doi     = {10.1063/1.438313},
}

@article{koster2003,
  author  = {K\"oster, Andreas M.},
  title   = {{Hermite} {Gaussian} auxiliary functions for the variational
             fitting of the {Coulomb} potential in density functional methods},
  journal = {J. Chem. Phys.},
  volume  = {118},
  pages   = {9943--9951},
  year    = {2003},
  doi     = {10.1063/1.1571519},
}

@article{tibshirani1996,
  title={Regression shrinkage and selection via the lasso},
  author={Tibshirani, Robert},
  journal={J. R. Statist. Soc. B},
  volume={58},
  pages={267--288},
  year={1996},
  doi={10.1111/j.2517-6161.1996.tb02080.x},
}

@article{bultinck2007b,
  author  = {Bultinck, Patrick and Ayers, Paul W. and Fias, Stijn and
             Tiels, Koen and Van Alsenoy, Christian},
  title   = {Uniqueness and basis set dependence of iterative {Hirshfeld}
             charges},
  journal = {Chem. Phys. Lett.},
  year    = {2007},
  doi     = {10.1016/j.cplett.2007.07.014},
}

@article{perdew1982,
  author  = {Perdew, John P. and Parr, Robert G. and Levy, Mel and
             Balduz, Jr., Jose L.},
  title   = {Density-Functional Theory for Fractional Particle Number:
             Derivative Discontinuities of the Energy},
  journal = {Phys. Rev. Lett.},
  volume  = {49},
  pages   = {1691--1694},
  year    = {1982},
  doi     = {10.1103/PhysRevLett.49.1691},
}

@article{nalewajski2000,
  author  = {Nalewajski, Roman F. and Parr, Robert G.},
  title   = {Information theory, atoms in molecules, and molecular
             similarity},
  journal = {Proc. Natl. Acad. Sci. U.S.A.},
  volume  = {97},
  pages   = {8879--8882},
  year    = {2000},
  doi     = {10.1073/pnas.97.16.8879},
}

@article{ayers2006,
  author  = {Ayers, Paul W.},
  title   = {Information theory, the shape function, and the {Hirshfeld}
             atom},
  journal = {Theor. Chem. Acc.},
  year    = {2006},
  doi     = {10.1007/s00214-006-0121-5},
}

@article{manz2019,
  author    = {Manz, Thomas A. and Chen, Taoyi and Cole, Daniel J. and Limas, Nidia Gabaldon and Fiszbein, Benjamin},
  title     = {New scaling relations to compute atom-in-material polarizabilities and dispersion coefficients: part 1. Theory and accuracy},
  journal   = {RSC Adv.},
  volume    = {9},
  pages     = {19297--19324},
  year      = {2019},
  doi       = {10.1039/c9ra03003d},
  url       = {https://doi.org/10.1039/c9ra03003d},
}

@article{rehman2026,
  title={Dispersion from polarizabilities of atoms in molecule: DPAIM},
  author={Rehman, Atta Ur and Shahbaz, Muhammad and Szalewicz, Krzysztof},
  journal={J. Chem. Theory Comput.},
  volume={22},
  number={16},
  pages={8368--8389},
  year={2026},
  publisher={ACS Publications},
  doi={10.1021/acs.jctc.6c01258},
}

@article{becke2005,
  author  = {Becke, Axel D. and Johnson, Erin R.},
  title   = {Exchange-hole dipole moment and the dispersion interaction},
  journal = {J. Chem. Phys.},
  volume  = {122},
  pages   = {154104},
  year    = {2005},
  doi     = {10.1063/1.1884601},
}

@article{becke2007,
  author  = {Becke, Axel D. and Johnson, Erin R.},
  title   = {Exchange-hole dipole moment and the dispersion interaction
             revisited},
  journal = {J. Chem. Phys.},
  volume  = {127},
  pages   = {154108},
  year    = {2007},
  doi     = {10.1063/1.2795701},
}

@article{becke1989,
  author  = {Becke, A. D. and Roussel, M. R.},
  title   = {Exchange holes in inhomogeneous systems: A coordinate-space
             model},
  journal = {Phys. Rev. A},
  volume  = {39},
  pages   = {3761--3767},
  year    = {1989},
  doi     = {10.1103/PhysRevA.39.3761},
}

@article{kirkwood1932,
  author  = {Kirkwood, J. G.},
  title   = {\"Uber die {Polarisierbarkeit} von {Atomen} und {Molek\"ulen}},
  journal = {Phys. Z.},
  volume  = {33},
  pages   = {57--60},
  year    = {1932},
}

@article{vinti1932,
  author  = {Vinti, John P.},
  title   = {A Relation Between the Electric and Diamagnetic Susceptibilities
             of Monatomic Gases},
  journal = {Phys. Rev.},
  volume  = {41},
  pages   = {813--817},
  year    = {1932},
  doi     = {10.1103/PhysRev.41.813},
}

@article{johnson2006,
  author  = {Johnson, Erin R. and Becke, Axel D.},
  title   = {A post-{Hartree}-{Fock} model of intermolecular interactions:
             Inclusion of higher-order corrections},
  journal = {J. Chem. Phys.},
  volume  = {124},
  pages   = {174104},
  year    = {2006},
  doi     = {10.1063/1.2190220},
}

@incollection{crc88,
  author    = {Miller, Thomas M.},
  title     = {Atomic and Molecular Polarizabilities},
  booktitle = {{CRC} Handbook of Chemistry and Physics},
  edition   = {88},
  editor    = {Lide, David R.},
  publisher = {CRC Press},
  address   = {Boca Raton, FL},
  year      = {2007},
  pages     = {10-193--10-202},
}

@article{xdmhybrid,
  author    = {Otero-de-la-Roza, A. and Johnson, Erin R.},
  title     = {Non-covalent interactions and thermochemistry using {XDM}-corrected hybrid and range-separated hybrid density functionals},
  journal   = {J. Chem. Phys.},
  volume    = {138},
  pages     = {204109},
  year      = {2013},
  doi       = {10.1063/1.4807330},
  url       = {https://doi.org/10.1063/1.4807330},
}

@article{becke1988b,
  author  = {Becke, A. D.},
  title   = {Density-functional exchange-energy approximation with correct
             asymptotic behavior},
  journal = {Phys. Rev. A},
  volume  = {38},
  pages   = {3098--3100},
  year    = {1988},
  doi     = {10.1103/PhysRevA.38.3098},
}

@article{lee1988,
  author  = {Lee, Chengteh and Yang, Weitao and Parr, Robert G.},
  title   = {Development of the {Colle-Salvetti} correlation-energy formula
             into a functional of the electron density},
  journal = {Phys. Rev. B},
  volume  = {37},
  pages   = {785--789},
  year    = {1988},
  doi     = {10.1103/PhysRevB.37.785},
}

@article{perdew1986,
  author  = {Perdew, John P.},
  title   = {Density-functional approximation for the correlation energy of
             the inhomogeneous electron gas},
  journal = {Phys. Rev. B},
  volume  = {33},
  pages   = {8822--8824},
  year    = {1986},
  doi     = {10.1103/PhysRevB.33.8822},
}

@article{perdew1986b,
  author  = {Perdew, John P. and Wang, Yue},
  title   = {Accurate and simple density functional for the electronic
             exchange energy: Generalized gradient approximation},
  journal = {Phys. Rev. B},
  volume  = {33},
  pages   = {8800--8802},
  year    = {1986},
  doi     = {10.1103/PhysRevB.33.8800},
}

@article{perdew1992,
  author  = {Perdew, John P. and Chevary, J. A. and Vosko, S. H. and
             Jackson, Koblar A. and Pederson, Mark R. and Singh, D. J. and
             Fiolhais, Carlos},
  title   = {Atoms, molecules, solids, and surfaces: Applications of the
             generalized gradient approximation for exchange and correlation},
  journal = {Phys. Rev. B},
  volume  = {46},
  pages   = {6671--6687},
  year    = {1992},
  doi     = {10.1103/PhysRevB.46.6671},
}

@article{becke1993,
  author  = {Becke, Axel D.},
  title   = {Density-functional thermochemistry. {III}. The role of exact
             exchange},
  journal = {J. Chem. Phys.},
  volume  = {98},
  pages   = {5648--5652},
  year    = {1993},
  doi     = {10.1063/1.464913},
}

@article{becke1993b,
  author  = {Becke, Axel D.},
  title   = {A new mixing of {Hartree-Fock} and local density-functional
             theories},
  journal = {J. Chem. Phys.},
  volume  = {98},
  pages   = {1372--1377},
  year    = {1993},
  doi     = {10.1063/1.464304},
}

@article{stephens1994,
  author  = {Stephens, P. J. and Devlin, F. J. and Chabalowski, C. F. and
             Frisch, M. J.},
  title   = {Ab Initio Calculation of Vibrational Absorption and Circular
             Dichroism Spectra Using Density Functional Force Fields},
  journal = {J. Phys. Chem.},
  volume  = {98},
  pages   = {11623--11627},
  year    = {1994},
  doi     = {10.1021/j100096a001},
}

@article{hamprecht1998,
  author  = {Hamprecht, Fred A. and Cohen, Aron J. and Tozer, David J. and
             Handy, Nicholas C.},
  title   = {Development and assessment of new exchange-correlation
             functionals},
  journal = {J. Chem. Phys.},
  volume  = {109},
  pages   = {6264--6271},
  year    = {1998},
  doi     = {10.1063/1.477267},
}

@article{adamo1999,
  author  = {Adamo, Carlo and Barone, Vincenzo},
  title   = {Toward reliable density functional methods without adjustable
             parameters: The {PBE0} model},
  journal = {J. Chem. Phys.},
  volume  = {110},
  pages   = {6158--6170},
  year    = {1999},
  doi     = {10.1063/1.478522},
}

@article{ernzerhof1999,
  author  = {Ernzerhof, Matthias and Scuseria, Gustavo E.},
  title   = {Assessment of the {Perdew-Burke-Ernzerhof} exchange-correlation
             functional},
  journal = {J. Chem. Phys.},
  volume  = {110},
  pages   = {5029--5036},
  year    = {1999},
  doi     = {10.1063/1.478401},
}

@article{yanai2004,
  author  = {Yanai, Takeshi and Tew, David P. and Handy, Nicholas C.},
  title   = {A new hybrid exchange--correlation functional using the
             {Coulomb}-attenuating method ({CAM-B3LYP})},
  journal = {Chem. Phys. Lett.},
  volume  = {393},
  pages   = {51--57},
  year    = {2004},
  doi     = {10.1016/j.cplett.2004.06.011},
}

@article{heyd2003,
  author  = {Heyd, Jochen and Scuseria, Gustavo E. and Ernzerhof, Matthias},
  title   = {Hybrid functionals based on a screened {Coulomb} potential},
  journal = {J. Chem. Phys.},
  volume  = {118},
  pages   = {8207--8215},
  year    = {2003},
  doi     = {10.1063/1.1564060},
}

@article{krukau2006,
  author  = {Krukau, Aliaksandr V. and Vydrov, Oleg A. and Izmaylov, Artur F.
             and Scuseria, Gustavo E.},
  title   = {Influence of the exchange screening parameter on the
             performance of screened hybrid functionals},
  journal = {J. Chem. Phys.},
  volume  = {125},
  pages   = {224106},
  year    = {2006},
  doi     = {10.1063/1.2404663},
}

@article{vydrov2006,
  author  = {Vydrov, Oleg A. and Scuseria, Gustavo E.},
  title   = {Assessment of a long-range corrected hybrid functional},
  journal = {J. Chem. Phys.},
  volume  = {125},
  pages   = {234109},
  year    = {2006},
  doi     = {10.1063/1.2409292},
}

@article{rohrdanz2009,
  author  = {Rohrdanz, Mary A. and Martins, Katie M. and Herbert, John M.},
  title   = {A long-range-corrected density functional that performs well for
             both ground-state properties and time-dependent density
             functional theory excitation energies, including charge-transfer
             excited states},
  journal = {J. Chem. Phys.},
  volume  = {130},
  pages   = {054112},
  year    = {2009},
  doi     = {10.1063/1.3073302},
}

@article{zhang1998,
  author  = {Zhang, Yingkai and Yang, Weitao},
  title   = {Comment on ``{Generalized} Gradient Approximation Made Simple''},
  journal = {Phys. Rev. Lett.},
  volume  = {80},
  pages   = {890},
  year    = {1998},
  doi     = {10.1103/PhysRevLett.80.890},
}

@article{tao2003,
  author  = {Tao, Jianmin and Perdew, John P. and Staroverov, Viktor N. and
             Scuseria, Gustavo E.},
  title   = {Climbing the Density Functional Ladder: Nonempirical
             Meta--Generalized Gradient Approximation Designed for Molecules
             and Solids},
  journal = {Phys. Rev. Lett.},
  volume  = {91},
  pages   = {146401},
  year    = {2003},
  doi     = {10.1103/PhysRevLett.91.146401},
}

@article{pslibrary,
  title={Pseudopotentials periodic table: From H to Pu},
  author={Dal Corso, Andrea},
  journal={Comput. Mater. Sci.},
  volume={95},
  pages={337--350},
  year={2014},
  doi={10.1016/j.commatsci.2014.07.043},
}

@article{paw,
  author={Bl{\"o}chl, P.E.},
  title={Projector augmented-wave method},
  journal={Phys. Rev. B},
  volume={50},
  year={1994},
  pages={17953},
  doi={10.1103/PhysRevB.50.17953},
}

@article{blum2009,
  author  = {Blum, Volker and Gehrke, Ralf and Hanke, Felix and Havu, Paula
             and Havu, Ville and Ren, Xinguo and Reuter, Karsten and
             Scheffler, Matthias},
  title   = {Ab initio molecular simulations with numeric atom-centered
             orbitals},
  journal = {Comput. Phys. Commun.},
  volume  = {180},
  pages   = {2175--2196},
  year    = {2009},
  doi     = {10.1016/j.cpc.2009.06.022},
}

@misc{g16,
  author       = {Frisch, M. J. and Trucks, G. W. and Schlegel, H. B. and
                  Scuseria, G. E. and Robb, M. A. and Cheeseman, J. R. and
                  others},
  title        = {Gaussian~16, Revision {A.03}},
  howpublished = {Gaussian, Inc., Wallingford, CT},
  year         = {2016},
}

@article{kannemann2010,
  author  = {Kannemann, Felix O. and Becke, Axel D.},
  title   = {van der {Waals} Interactions in Density-Functional Theory:
             Intermolecular Complexes},
  journal = {J. Chem. Theory Comput.},
  volume  = {6},
  pages   = {1081--1088},
  year    = {2010},
  doi     = {10.1021/ct900699r},
}

@article{price2023,
  author  = {Price, Alastair J. A. and Otero-de-la-Roza, Alberto and
             Johnson, Erin R.},
  title   = {{XDM}-corrected hybrid {DFT} with numerical atomic orbitals
             predicts molecular crystal lattice energies with unprecedented
             accuracy},
  journal = {Chem. Sci.},
  volume  = {14},
  pages   = {1252--1262},
  year    = {2023},
  doi     = {10.1039/d2sc05997e},
}

@article{schwerdtfeger2019,
  author  = {Schwerdtfeger, Peter and Nagle, Jeffrey K.},
  title   = {2018 Table of Static Dipole Polarizabilities of the Neutral
             Elements in the Periodic Table},
  journal = {Mol. Phys.}, volume = {117}, pages = {1200--1225}, year = {2019},
  doi     = {10.1080/00268976.2018.1535143},
}

@article{schwerdtfeger2019b,
  author  = {Schwerdtfeger, Peter and Nagle, Jeffrey K.},
  title   = {Correction to: 2018 Table of Static Dipole Polarizabilities of
             the Neutral Elements in the Periodic Table},
  journal = {Mol. Phys.}, volume = {117}, pages = {1585}, year = {2019},
  doi     = {10.1080/00268976.2018.1549647},
}

@article{wilkins2019,
  author  = {Wilkins, David M. and Grisafi, Andrea and Yang, Yang and
             Lao, Ka Un and {DiStasio, Jr.}, Robert A. and Ceriotti, Michele},
  title   = {Accurate molecular polarizabilities with coupled cluster theory
             and machine learning},
  journal = {Proc. Natl. Acad. Sci. U.S.A.}, volume = {116},
  pages   = {3401--3406}, year = {2019},
  doi     = {10.1073/pnas.1816132116},
}

@article{yang2019,
  author  = {Yang, Yang and Lao, Ka Un and Wilkins, David M. and
             Grisafi, Andrea and Ceriotti, Michele and
             {DiStasio, Jr.}, Robert A.},
  title   = {Quantum mechanical static dipole polarizabilities in the {QM7b}
             and {AlphaML} showcase databases},
  journal = {Sci. Data}, volume = {6}, pages = {152}, year = {2019},
  doi     = {10.1038/s41597-019-0157-8},
}

@article{tkatchenko2012,
  author  = {Tkatchenko, Alexandre and DiStasio, Jr., Robert A. and
             Car, Roberto and Scheffler, Matthias},
  title   = {Accurate and Efficient Method for Many-Body van der {Waals}
             Interactions},
  journal = {Phys. Rev. Lett.}, volume = {108}, pages = {236402}, year = {2012},
  doi     = {10.1103/PhysRevLett.108.236402},
}

@article{grimme2011b,
  author  = {Grimme, Stefan},
  title   = {Density functional theory with {London} dispersion corrections},
  journal = {WIREs Comput. Mol. Sci.}, volume = {1}, pages = {211--228},
  year    = {2011}, doi = {10.1002/wcms.30},
}

@article{caldeweyher2019,
  author  = {Caldeweyher, Eike and Ehlert, Sebastian and Hansen, Andreas and
             Neugebauer, Hagen and Spicher, Sebastian and Bannwarth, Christoph
             and Grimme, Stefan},
  title   = {A generally applicable atomic-charge dependent {London}
             dispersion correction},
  journal = {J. Chem. Phys.}, volume = {150}, pages = {154122}, year = {2019},
  doi     = {10.1063/1.5090222},
}

@article{kumar1985,
  author={Kumar, Ashok and Meath, William J.},
  title={Pseudo-spectral dipole oscillator strengths and dipole-dipole and triple-dipole dispersion energy coefficients for {HF}, {HCl}, {HBr}, {He}, {Ne}, {Ar}, {Kr} and {Xe}},
  journal={Mol. Phys.},
  volume={54},
  pages={823--833},
  year={1985},
  doi={10.1080/00268978500103191},
}

@article{montavon2013,
  author  = {Montavon, Gr{\'e}goire and Rupp, Matthias and Gobre, Vivekanand
             and Vazquez-Mayagoitia, Alvaro and Hansen, Katja and
             Tkatchenko, Alexandre and M{\"u}ller, Klaus-Robert and
             {von Lilienfeld}, O. Anatole},
  title   = {Machine learning of molecular electronic properties in chemical
             compound space},
  journal = {New J. Phys.},
  volume  = {15},
  pages   = {095003},
  year    = {2013},
  doi     = {10.1088/1367-2630/15/9/095003},
}

@article{rupp2012,
  author  = {Rupp, Matthias and Tkatchenko, Alexandre and
             M{\"u}ller, Klaus-Robert and {von Lilienfeld}, O. Anatole},
  title   = {Fast and Accurate Modeling of Molecular Atomization Energies
             with Machine Learning},
  journal = {Phys. Rev. Lett.},
  volume  = {108},
  pages   = {058301},
  year    = {2012},
  doi     = {10.1103/PhysRevLett.108.058301},
}

@article{blum2009b,
  author  = {Blum, Lorenz C. and Reymond, Jean-Louis},
  title   = {970 Million Druglike Small Molecules for Virtual Screening in
             the Chemical Universe Database {GDB-13}},
  journal = {J. Am. Chem. Soc.},
  volume  = {131},
  pages   = {8732--8733},
  year    = {2009},
  doi     = {10.1021/ja902302h},
}

@misc{yang2019b,
  author       = {Yang, Yang and Lao, Ka Un and Wilkins, David M. and
                  Grisafi, Andrea and Ceriotti, Michele and
                  {DiStasio, Jr.}, Robert A.},
  title        = {Quantum mechanical static dipole polarizabilities in the
                  {QM7b} and {AlphaML} showcase databases},
  howpublished = {Materials Cloud Archive},
  year         = {2019},
  doi          = {10.24435/materialscloud:2019.0002},
  url          = {https://doi.org/10.24435/materialscloud:2019.0002},
}

@article{jurecka2006,
  author  = {Jure{\v{c}}ka, Petr and {\v{S}}poner, Ji{\v{r}}{\'\i} and
             {\v{C}}ern{\'y}, Ji{\v{r}}{\'\i} and Hobza, Pavel},
  title   = {Benchmark database of accurate ({MP2} and {CCSD(T)} complete
             basis set limit) interaction energies of small model complexes,
             {DNA} base pairs, and amino acid pairs},
  journal = {Phys. Chem. Chem. Phys.}, volume = {8}, pages = {1985--1993},
  year    = {2006}, doi = {10.1039/b600027d},
}

@article{grafova2010,
  author  = {Gr{\'a}fov{\'a}, Lucie and Pito{\v{n}}{\'a}k, Michal and
             {\v{R}}ez{\'a}{\v{c}}, Jan and Hobza, Pavel},
  title   = {Comparative Study of Selected Wave Function and Density
             Functional Methods for Noncovalent Interaction Energy
             Calculations Using the Extended {S22} Data Set},
  journal = {J. Chem. Theory Comput.}, volume = {6}, pages = {2365--2376},
  year    = {2010}, doi = {10.1021/ct1002253},
}

@article{rezac2011,
  author  = {{\v{R}}ez{\'a}{\v{c}}, Jan and Riley, Kevin E. and Hobza, Pavel},
  title   = {{S66}: A Well-balanced Database of Benchmark Interaction Energies
             Relevant to Biomolecular Structures},
  journal = {J. Chem. Theory Comput.}, volume = {7}, pages = {2427--2438},
  year    = {2011}, doi = {10.1021/ct2002946},
}

@article{brauer2016,
  author  = {Brauer, Brina and Kesharwani, Manoj K. and Kozuch, Sebastian and
             Martin, Jan M. L.},
  title   = {The {S66x8} benchmark for noncovalent interactions revisited:
             explicitly correlated ab initio methods and density functional
             theory},
  journal = {Phys. Chem. Chem. Phys.}, volume = {18}, pages = {20905--20925},
  year    = {2016}, doi = {10.1039/c6cp00688d},
}

@article{rezac2020,
  author  = {{\v{R}}ez{\'a}{\v{c}}, Jan},
  title   = {Non-Covalent Interactions Atlas Benchmark Data Sets: Hydrogen
             Bonding},
  journal = {J. Chem. Theory Comput.}, volume = {16}, pages = {2355--2368},
  year    = {2020}, doi = {10.1021/acs.jctc.9b01265},
}

@article{kriz2022,
  author  = {K{\v{r}}{\'\i}{\v{z}}, Kristian and {\v{R}}ez{\'a}{\v{c}}, Jan},
  title   = {Non-covalent interactions atlas benchmark data sets 4:
             $\sigma$-hole interactions},
  journal = {Phys. Chem. Chem. Phys.}, volume = {24}, pages = {14794--14804},
  year    = {2022}, doi = {10.1039/d2cp01600a},
}

@article{spicher2021,
  author  = {Spicher, Sebastian and Caldeweyher, Eike and Hansen, Andreas and
             Grimme, Stefan},
  title   = {Benchmarking London dispersion corrected density functional
             theory for noncovalent ion--$\pi$ interactions},
  journal = {Phys. Chem. Chem. Phys.}, volume = {23}, pages = {11635--11648},
  year    = {2021}, doi = {10.1039/d1cp01333e},
}

@article{goerigk2017,
  author  = {Goerigk, Lars and Hansen, Andreas and Bauer, Christoph and
             Ehrlich, Stephan and Najibi, Asim and Grimme, Stefan},
  title   = {A look at the density functional theory zoo with the advanced
             {GMTKN55} database for general main group thermochemistry,
             kinetics and noncovalent interactions},
  journal = {Phys. Chem. Chem. Phys.},
  volume  = {19},
  pages   = {32184--32215},
  year    = {2017},
  doi     = {10.1039/c7cp04913g},
}

@article{nickerson2023,
  author  = {Nickerson, Cameron J. and Bryenton, Kyle R. and
             Price, Alastair J. A. and Johnson, Erin R.},
  title   = {Comparison of Density-Functional Theory Dispersion Corrections
             for the {DES15K} Database},
  journal = {J. Phys. Chem. A},
  volume  = {127},
  pages   = {8712--8722},
  year    = {2023},
  doi     = {10.1021/acs.jpca.3c04332},
}

@article{caldeweyher2020,
  author  = {Caldeweyher, Eike and Mewes, Jan-Michael and Ehlert, Sebastian
             and Grimme, Stefan},
  title   = {Extension and evaluation of the {D4} London-dispersion model for
             periodic systems},
  journal = {Phys. Chem. Chem. Phys.},
  volume  = {22},
  pages   = {8499--8512},
  year    = {2020},
  doi     = {10.1039/d0cp00502a},
}

@article{grimme2010,
  author  = {Grimme, Stefan and Antony, Jens and Ehrlich, Stephan and
             Krieg, Helge},
  title   = {A consistent and accurate \emph{ab initio} parametrization of
             density functional dispersion correction ({DFT-D}) for the 94
             elements {H-Pu}},
  journal = {J. Chem. Phys.},
  volume  = {132},
  pages   = {154104},
  year    = {2010},
  doi     = {10.1063/1.3382344},
}

@article{ambrosetti2014,
  author  = {Ambrosetti, Alberto and Reilly, Anthony M. and
             {DiStasio, Jr.}, Robert A. and Tkatchenko, Alexandre},
  title   = {Long-range correlation energy calculated from coupled atomic
             response functions},
  journal = {J. Chem. Phys.},
  volume  = {140},
  pages   = {18A508},
  year    = {2014},
  doi     = {10.1063/1.4865104},
}

@article{steinmann2011,
  author  = {Steinmann, Stephan N. and Corminboeuf, Cl{\'e}mence},
  title   = {A generalized-gradient approximation exchange hole model for
             dispersion coefficients},
  journal = {J. Chem. Phys.},
  volume  = {134},
  pages   = {044117},
  year    = {2011},
  doi     = {10.1063/1.3545985},
}

@article{tessman1953,
  author  = {Tessman, Jack R. and Kahn, A. H. and Shockley, William},
  title   = {Electronic Polarizabilities of Ions in Crystals},
  journal = {Phys. Rev.},
  volume  = {92},
  pages   = {890--895},
  year    = {1953},
  doi     = {10.1103/PhysRev.92.890},
}

@article{oterodelaroza2020,
  author  = {{Otero-de-la-Roza}, A. and Johnson, Erin R.},
  title   = {Application of {XDM} to ionic solids: The importance of
             dispersion for bulk moduli and crystal geometries},
  journal = {J. Chem. Phys.},
  volume  = {153},
  pages   = {054121},
  year    = {2020},
  doi     = {10.1063/5.0015133},
}

@article{doll1997,
  author  = {Doll, Klaus and Stoll, Hermann},
  title   = {Cohesive properties of alkali halides},
  journal = {Phys. Rev. B},
  volume  = {56},
  pages   = {10121--10127},
  year    = {1997},
  doi     = {10.1103/PhysRevB.56.10121},
}

@article{doll1998,
  author  = {Doll, Klaus and Stoll, Hermann},
  title   = {Ground-state properties of heavy alkali halides},
  journal = {Phys. Rev. B},
  volume  = {57},
  pages   = {4327--4331},
  year    = {1998},
  doi     = {10.1103/PhysRevB.57.4327},
}

@article{bjorkman2014,
  author  = {Bj{\"o}rkman, Torbj{\"o}rn},
  title   = {Testing several recent van der {W}aals density functionals for
             layered structures},
  journal = {J. Chem. Phys.},
  volume  = {141},
  pages   = {074708},
  year    = {2014},
  doi     = {10.1063/1.4893329},
}

@article{hummel2016,
  author  = {Hummel, Felix and Gruber, Thomas and Gr{\"u}neis, Andreas},
  title   = {A many-electron perturbation theory study of the hexagonal boron
             nitride bilayer system},
  journal = {Eur. Phys. J. B},
  volume  = {89},
  pages   = {235},
  year    = {2016},
  doi     = {10.1140/epjb/e2016-70177-4},
}

@article{oterodelaroza2020b,
  author  = {{Otero-de-la-Roza}, A. and LeBlanc, Luc M. and Johnson, Erin R.},
  title   = {Asymptotic Pairwise Dispersion Corrections Can Describe Layered
             Materials Accurately},
  journal = {J. Phys. Chem. Lett.},
  volume  = {11},
  pages   = {2298--2302},
  year    = {2020},
  doi     = {10.1021/acs.jpclett.0c00348},
}

@article{rumson2026,
  author  = {Rumson, Adrian F. and Bryenton, Kyle R. and Johnson, Erin R.},
  title   = {The effects of dispersion damping and three-body interactions for
             accurate layered-material exfoliation energies},
  journal = {Phys. Chem. Chem. Phys.},
  volume  = {28},
  pages   = {13543--13552},
  year    = {2026},
  doi     = {10.1039/d6cp01304j},
}

@article{oterodelaroza2012,
  author  = {{Otero-de-la-Roza}, A. and Johnson, Erin R.},
  title   = {A benchmark for non-covalent interactions in solids},
  journal = {J. Chem. Phys.},
  volume  = {137},
  pages   = {054103},
  year    = {2012},
  doi     = {10.1063/1.4738961},
}

@article{reilly2013,
  author  = {Reilly, Anthony M. and Tkatchenko, Alexandre},
  title   = {Understanding the role of vibrations, exact exchange, and
             many-body van der {W}aals interactions in the cohesive properties
             of molecular crystals},
  journal = {J. Chem. Phys.},
  volume  = {139},
  pages   = {024705},
  year    = {2013},
  doi     = {10.1063/1.4812819},
}

@article{dolgonos2019,
  author  = {Dolgonos, Grygoriy A. and Hoja, Johannes and Boese, A. Daniel},
  title   = {Revised values for the {X23} benchmark set of molecular crystals},
  journal = {Phys. Chem. Chem. Phys.},
  volume  = {21},
  pages   = {24333--24344},
  year    = {2019},
  doi     = {10.1039/c9cp04488d},
}

@article{brandenburg2015,
  author  = {Brandenburg, Jan Gerit and Maas, Tilo and Grimme, Stefan},
  title   = {Benchmarking {DFT} and semiempirical methods on structures and
             lattice energies for ten ice polymorphs},
  journal = {J. Chem. Phys.},
  volume  = {142},
  pages   = {124104},
  year    = {2015},
  doi     = {10.1063/1.4916070},
}

@article{dellapia2022,
  author  = {{Della Pia}, Flaviano and Zen, Andrea and Alf{\`e}, Dario and
             Michaelides, Angelos},
  title   = {{DMC-ICE13}: Ambient and high pressure polymorphs of ice from
             diffusion Monte Carlo and density functional theory},
  journal = {J. Chem. Phys.},
  volume  = {157},
  pages   = {134701},
  year    = {2022},
  doi     = {10.1063/5.0102645},
}

@article{tao2017,
  author  = {Tao, Jianmin and Zheng, Fan and Gebhardt, Julian and
             Perdew, John P. and Rappe, Andrew M.},
  title   = {Screened van der {W}aals correction to density functional theory
             for solids},
  journal = {Phys. Rev. Mater.},
  volume  = {1},
  pages   = {020802(R)},
  year    = {2017},
  doi     = {10.1103/PhysRevMaterials.1.020802},
}

@book{sirdeshmukh2001,
  author    = {Sirdeshmukh, D. B. and Sirdeshmukh, L. and Subhadra, K. G.},
  title     = {Alkali Halides: A Handbook of Physical Properties},
  series    = {Springer Series in Materials Science},
  volume    = {49},
  publisher = {Springer},
  address   = {Berlin},
  year      = {2001},
  doi       = {10.1007/978-3-662-04341-7},
}

@article{gibbs2a,
  author={{Otero-de-la-Roza}, A. and {Luaña}, V.},
  title={{G}ibbs2: a new version of the quasi-harmonic model code. {I}. {R}obust treatment of the static data},
  journal={Comput. Phys. Commun.},
  volume={182},
  pages={1708--1720},
  year={2011},
}

@article{gibbs2b,
  author={{Otero-de-la-Roza}, A. and {Abbasi-Pérez}, D. and {Luaña}, V.},
  title={{G}ibbs2: a new version of the quasi-harmonic model code. {II}. {M}odels for solid-state thermodynamics, features and implementation.},
  journal={Comput. Phys. Commun.},
  volume={182},
  pages={2232--2248},
  year={2011},
}

@article{szabo2022,
  author    = {Szab{\'o}, P{\'e}ter and G{\'o}ger, Szabolcs and Charry, Jorge and Karimpour, Mohammad Reza and Fedorov, Dmitry V. and Tkatchenko, Alexandre},
  title     = {Four-Dimensional Scaling of Dipole Polarizability in Quantum Systems},
  journal   = {Phys. Rev. Lett.},
  volume    = {128},
  pages     = {070602},
  year      = {2022},
  doi       = {10.1103/physrevlett.128.070602},
  url       = {https://doi.org/10.1103/physrevlett.128.070602},
}

@article{goger2024,
  author    = {G{\'o}ger, Szabolcs and Karimpour, Mohammad Reza and Tkatchenko, Alexandre},
  title     = {Four-Dimensional Scaling of Dipole Polarizability: From Single-Particle Models to Atoms and Molecules},
  journal   = {J. Chem. Theory Comput.},
  volume    = {20},
  pages     = {6621--6631},
  year      = {2024},
  doi       = {10.1021/acs.jctc.4c00582},
  url       = {https://doi.org/10.1021/acs.jctc.4c00582},
}

@article{stohr2019,
  author    = {St{\"o}hr, Martin and Van Voorhis, Troy and Tkatchenko, Alexandre},
  title     = {Theory and practice of modeling {van der Waals} interactions in electronic-structure calculations},
  journal   = {Chem. Soc. Rev.},
  volume    = {48},
  pages     = {4118--4154},
  year      = {2019},
  doi       = {10.1039/c9cs00060g},
  url       = {https://doi.org/10.1039/c9cs00060g},
}

@article{beran2016,
  author    = {Beran, Gregory J. O.},
  title     = {Modeling Polymorphic Molecular Crystals with Electronic Structure Theory},
  journal   = {Chem. Rev.},
  volume    = {116},
  pages     = {5567--5613},
  year      = {2016},
  doi       = {10.1021/acs.chemrev.5b00648},
  url       = {https://doi.org/10.1021/acs.chemrev.5b00648},
}

@article{hoja2019,
  author    = {Hoja, Johannes and Ko, Hsin-Yu and Neumann, Marcus A. and Car, Roberto and DiStasio, Robert A. and Tkatchenko, Alexandre},
  title     = {Reliable and practical computational description of molecular crystal polymorphs},
  journal   = {Sci. Adv.},
  volume    = {5},
  pages     = {eaau3338},
  year      = {2019},
  doi       = {10.1126/sciadv.aau3338},
  url       = {https://doi.org/10.1126/sciadv.aau3338},
}

@article{grimme2016,
  author    = {Grimme, Stefan and Hansen, Andreas and Brandenburg, Jan Gerit and Bannwarth, Christoph},
  title     = {Dispersion-Corrected Mean-Field Electronic Structure Methods},
  journal   = {Chem. Rev.},
  volume    = {116},
  pages     = {5105--5154},
  year      = {2016},
  doi       = {10.1021/acs.chemrev.5b00533},
  url       = {https://doi.org/10.1021/acs.chemrev.5b00533},
}

@article{hermann2017,
  author    = {Hermann, Jan and DiStasio, Robert A. and Tkatchenko, Alexandre},
  title     = {First-Principles Models for {van der Waals} Interactions in Molecules and Materials: Concepts, Theory, and Applications},
  journal   = {Chem. Rev.},
  volume    = {117},
  pages     = {4714--4758},
  year      = {2017},
  doi       = {10.1021/acs.chemrev.6b00446},
  url       = {https://doi.org/10.1021/acs.chemrev.6b00446},
}

@article{dion2004,
  author    = {Dion, M. and Rydberg, H. and Schr{\"o}der, E. and Langreth, D. C. and Lundqvist, B. I.},
  title     = {Van der Waals Density Functional for General Geometries},
  journal   = {Phys. Rev. Lett.},
  volume    = {92},
  pages     = {246401},
  year      = {2004},
  doi       = {10.1103/PhysRevLett.92.246401},
  url       = {https://doi.org/10.1103/PhysRevLett.92.246401},
}

@article{vydrov2010b,
  author    = {Vydrov, Oleg A. and Van Voorhis, Troy},
  title     = {Nonlocal {van der Waals} density functional: The simpler the better},
  journal   = {J. Chem. Phys.},
  volume    = {133},
  pages     = {244103},
  year      = {2010},
  doi       = {10.1063/1.3521275},
  url       = {https://doi.org/10.1063/1.3521275},
}

@incollection{johnson2017,
  author    = {Johnson, Erin R.},
  title     = {The Exchange-Hole Dipole Moment Dispersion Model},
  booktitle = {Non-Covalent Interactions in Quantum Chemistry and Physics},
  editor    = {Otero-de-la-Roza, Alberto and DiLabio, Gino A.},
  publisher = {Elsevier},
  pages     = {169--194},
  year      = {2017},
  doi       = {10.1016/b978-0-12-809835-6.00006-2},
  url       = {https://doi.org/10.1016/b978-0-12-809835-6.00006-2},
}

@article{price2023b,
  author    = {Price, Alastair J. A. and Mayo, R. Alex and Otero-de-la-Roza, Alberto and Johnson, Erin R.},
  title     = {Accurate and efficient polymorph energy ranking with {XDM}-corrected hybrid {DFT}},
  journal   = {CrystEngComm},
  volume    = {25},
  pages     = {953--960},
  year      = {2023},
  doi       = {10.1039/d2ce01594c},
  url       = {https://doi.org/10.1039/d2ce01594c},
}

@article{mayo2024,
  author    = {Mayo, R. Alex and Price, Alastair J. A. and Otero-de-la-Roza, Alberto and Johnson, Erin R.},
  title     = {Assessment of the exchange-hole dipole moment dispersion correction for the energy ranking stage of the seventh crystal structure prediction blind test},
  journal   = {Acta Crystallogr. B},
  volume    = {80},
  pages     = {595--605},
  year      = {2024},
  doi       = {10.1107/s2052520624002774},
  url       = {https://doi.org/10.1107/s2052520624002774},
}

@article{christian2016,
  author    = {Christian, Matthew S. and Otero-de-la-Roza, Alberto and Johnson, Erin R.},
  title     = {Surface Adsorption from the Exchange-Hole Dipole Moment Dispersion Model},
  journal   = {J. Chem. Theory Comput.},
  volume    = {12},
  pages     = {3305--3315},
  year      = {2016},
  doi       = {10.1021/acs.jctc.6b00222},
  url       = {https://doi.org/10.1021/acs.jctc.6b00222},
}

@article{parr2005,
  author    = {Parr, Robert G. and Ayers, Paul W. and Nalewajski, Roman F.},
  title     = {What Is an Atom in a Molecule?},
  journal   = {J. Phys. Chem. A},
  volume    = {109},
  pages     = {3957--3959},
  year      = {2005},
  doi       = {10.1021/jp0404596},
  url       = {https://doi.org/10.1021/jp0404596},
}

@article{ayers2000,
  author    = {Ayers, Paul W.},
  title     = {Atoms in molecules, an axiomatic approach. {I}. Maximum transferability},
  journal   = {J. Chem. Phys.},
  volume    = {113},
  pages     = {10886--10898},
  year      = {2000},
  doi       = {10.1063/1.1327268},
  url       = {https://doi.org/10.1063/1.1327268},
}

@article{kim2020,
  author    = {Kim, Minho and Kim, Won June and Gould, Timothy and Lee, Eok Kyun and Leb{\`e}gue, S{\'e}bastien and Kim, Hyungjun},
  title     = {uMBD: A Materials-Ready Dispersion Correction That Uniformly Treats Metallic, Ionic, and {van der Waals} Bonding},
  journal   = {J. Am. Chem. Soc.},
  volume    = {142},
  pages     = {2346--2354},
  year      = {2020},
  doi       = {10.1021/jacs.9b11589},
  url       = {https://doi.org/10.1021/jacs.9b11589},
}

@article{buckingham1937,
  author    = {Buckingham, R. A.},
  title     = {The quantum theory of atomic polarization I--- Polarization by a uniform field},
  journal   = {Proc. R. Soc. A},
  volume    = {160},
  pages     = {94--113},
  year      = {1937},
  doi       = {10.1098/rspa.1937.0097},
  url       = {https://doi.org/10.1098/rspa.1937.0097},
}

@article{xdmbasis,
  author={{Johnson}, E. R. and {Otero-de-la-Roza}, A. and {Dale}, S. G. and {DiLabio}, G. A.},
  title={Efficient basis sets for non-covalent interactions in density-functional theory},
  journal={J. Chem. Phys.},
  volume={139},
  pages={214109},
  year={2013},
}

@article{xdmc9,
  author={{Otero-de-la-Roza}, A. and {Johnson}, E. R.},
  title={Many-body dispersion interactions from the exchange-hole dipole moment model},
  journal={J. Chem. Phys.},
  volume={138},
  pages={054103},
  year={2013},
}

@article{laturia2018,
  author    = {Laturia, Akash and Van de Put, Maarten L. and Vandenberghe, William G.},
  title     = {Dielectric properties of hexagonal boron nitride and transition metal dichalcogenides: from monolayer to bulk},
  journal   = {npj 2D Mater. Appl.},
  volume    = {2},
  pages     = {6},
  year      = {2018},
  doi       = {10.1038/s41699-018-0050-x},
  url       = {https://doi.org/10.1038/s41699-018-0050-x},
}

\end{document}